\documentclass[usenatbib]{mn2e}
\usepackage{epsf,graphics,graphicx}
\usepackage{amsmath}
\usepackage{amssymb,latexsym,mathrsfs, bm}
\usepackage{color}
\usepackage{float}
\usepackage{natbib}
\usepackage{times}
\usepackage{pdflscape}
\usepackage{longtable}

\def\eq#1{{Eq.~(\ref{#1})}}

\title[HI intensity fluctuations]{Theoretical and observational constraints on the
HI intensity power spectrum}
\author[Padmanabhan, Choudhury and Refregier]{Hamsa
Padmanabhan$^{1}$\thanks{Electronic address: hamsa@iucaa.ernet.in},
T. Roy Choudhury$^2$\thanks{Electronic address:
{tirth@ncra.tifr.res.in}},
 Alexandre Refregier$^3$\thanks{Electronic address: {alexandre.refregier@phys.ethz.ch}
}\\
$^{1}$ Inter-University Centre for Astronomy and Astrophysics, Pune 411007,
India\\
$^{2}$ National Centre for Radio Astrophysics, Tata Institute of Fundamental
Research, Pune 411007, India \\
$^{3}$  Institute for Astronomy, Eidgen\"{o}ssische Technische Hochschule Zurich, Wolfgang-Pauli-Strasse 27, CH-8093 Z\"{u}rich, Switzerland}

\begin{document}
\date{ }
\maketitle

\begin{abstract}
Mapping of the neutral hydrogen (HI) 21-cm intensity fluctuations across redshifts promises a novel and powerful probe of cosmology. The neutral hydrogen gas mass density, $\Omega_{\rm HI}$ and bias parameter, $b_{\rm HI}$ are key astrophysical inputs to the HI intensity fluctuation power spectrum. We compile the latest theoretical and observational constraints on  $\Omega_{\rm HI}$ and $b_{\rm HI}$ at various redshifts in the post-reionization universe. Constraints are incorporated from galaxy surveys, HI intensity mapping experiments, damped Lyman-$\alpha$ system observations, theoretical prescriptions for assigning HI to dark matter halos, and the results of numerical simulations. {{Using a minimum variance interpolation scheme, we obtain the predicted uncertainties on the HI intensity fluctuation power spectrum across redshifts 0-3.5 for three different confidence scenarios.}} We provide a convenient tabular form for the interpolated values of $\Omega_{\rm HI}$,
$b_{\rm HI}$ and the HI power spectrum amplitude and their uncertainties. We discuss the consequences for the measurement of the power spectrum by current and future intensity mapping experiments.
\end{abstract}


\section{Introduction}
 
Since the theoretical predictions by Hendrik van der Hulst in 1944 and the first observations by \citet{ewenpurcell1951} and \citet{mulleroort1951}, the 21 cm hyperfine line of hydrogen remains a powerful probe of the HI content of galaxies and now promises to revolutionize observational cosmology. This emission line allows for the measurement of the intensity of fluctuations across frequency ranges or equivalently across cosmic time, thus making it a three-dimensional probe of the universe. It promises to probe a much larger comoving volume than galaxy surveys in the visible band, and consequently may lead to higher precision in the measurement of the matter power spectrum and cosmological parameters. Since the power spectrum extends to the Jeans' length of the baryonic material, it allows sensitivity to much smaller scales than probed by the CMB. The inherent weakness of the line transition prevents the saturation of the line, thus enabling it to serve as a direct probe of the neutral gas content of the intergalactic medium during the dark 
ages and cosmic dawn prior to the 
epoch of hydrogen reionization.

{In the post-reionization epoch ($z \lesssim 6$), the 21-cm line emission is expected to provide a tracer of the underlying dark matter distribution due to the absence of the complicated reionization astrophysics, hence it may be used to study the large-scale structure at intermediate redshifts \citep{bharadwaj2001a, bharadwaj2001, bharadwaj2004, wyithe2008, wyithe2009, bharadwaj2009, wyithe2010}. HI gas in galaxies and their environments is also a tool to understand the physics of galaxy evolution \citep{wyithe2008b}. At low redshifts $z \sim 1$, these observations are also expected to serve as a useful probe of dark energy  \citep{chang10}; the acoustic oscillations in the power spectrum may be used to constrain dark energy out to high redshifts $z \gtrsim 3.5$ \citep{wyithe2008a}.} 
 
 Several surveys, both ongoing and being planned for the future, aim to observe and map the neutral hydrogen content in the local and high-redshift universe. These include the HI Parkes All-Sky Survey \citep[][HIPASS]{barnes2001, meyer2004, zwaan05}, the HI Jodrell All-Sky Survey \citep[][HIJASS]{lang2003}, the Blind Ultra-Deep HI Environmental Survey \citep[][BUDHIES]{jaffe2012} which searches for HI in galaxy cluster environments with the Westerbork Synthesis Radio Telescope (WSRT),\footnote{http://www.astron.nl/radio-observatory/astronomers/wsrt-astronomers} with other surveys using the WSRT presenting complementary measurements of HI content in field galaxies \citep{rhee13}. Other current surveys include the Arecibo Fast Legacy ALFA Survey \citep[][ALFALFA]{giovanelli2005, martin10} and the GALEX Arecibo SDSS Survey (GASS) which measures the HI intensity fluctuations on $\sim 1000$ optically selected galaxies \citep{catinella2010} over the redshift interval $0.025 < z < 0.05$. The Giant Meterwave Radio Telescope \citep[][GMRT]{swarup1991} may be used to map the 21-cm diffuse background out to $z \sim 0.4$ by signal stacking measurements \citep{lah07, lah2009}. The Ooty Radio Telescope (ORT)\footnote{http://rac.ncra.tifr.res.in} may also be used to map the HI intensity fluctuation at redshift 3.35 \citep{saiyadali2013}. Future experiments, with telescopes under development, include the Murchinson Widefield Array (MWA),\footnote{http://www.mwatelescope.org} the Square Kilometre Array (SKA),\footnote{https://www.skatelescope.org} the Low Frequency Array (LOFAR),\footnote{http://www.lofar.org} the Precision Array to Probe the Epoch of Reionization (PAPER),\footnote{http://eor.berkeley.edu} the WSRT APERture Tile In Focus (APERTIF) survey \citep{oosterloo2010}, the Karl G. Jansky Very Large Array (JVLA),\footnote{https://science.nrao.edu/facilities/vla} the Meer-Karoo Array Telescope \citep[][MeerKAT]{jonas2009} 
and the Australian SKA Pathfinder \citep[][ASKAP]{johnston2008} Wallaby Survey. Many of these telescopes will map the neutral hydrogen content at higher redshifts, $z \sim 6-50$ as well. 
 
 There are also surveys that map the HI 21-cm intensity of the universe at intermediate redshifts without the detection of individual galaxies. A three-dimensional intensity map of 21-cm emission at $z \sim 0.53-1.12$ has been presented in \citet{chang10} using the Green Bank Telescope (GBT). The Effelsberg-Bonn survey is an all-sky survey having covered 8000 deg$^2$ out to redshift 0.07 \citep{kerp2011}. Several intensity mapping experiments  over redshifts $z \sim 0.5 - 2.5$, including the Baryon Acoustic Oscillation Broadband and Broad-beam \citep[][BAOBAB]{pober2013}, BAORadio \citep{ansari2012}, BAO from Integrated Neutral Gas Observations \citep[][BINGO]{battye2012}, CHIME\footnote{http://chime.phas.ubc.ca} and TianLai \citep{chen2012} are being planned for the future. 
At high redshifts, $z \sim 1.5 - 5$, the current major observational probes of the neutral hydrogen content have been Damped Lyman Alpha absorption systems (DLAs). The latest surveys of DLAs include those from the HST and the SDSS \citep{rao06, prochaska09, noterdaeme09, noterdaeme12} and the ESO/UVES \citep{zafar2013} which trace the HI content in and around galaxies in the spectra of high redshift background quasars. The bias parameter for DLAs has been recently measured in the Baryon Oscillation Spectroscopic Survey (BOSS) by estimating their cross-correlation with the Lyman-$\alpha$ forest \citep{fontribera2012} and leads to the computation of the DLA bias at redshift 2.3.

On the theoretical front, cosmological hydrodynamical simulations have been used to investigate the neutral hydrogen content of the post-reionization universe \citep{duffy2012, rahmati2013, dave2013} using detailed modelling of self-shielding, galactic outflows and radiative transfer. The simulations have been found to produce results that match the observed neutral hydrogen fractions and column densities for physically motivated models of star formation and outflows. Analytical prescriptions for assigning HI to halos have also been used to model the bias parameter of HI-selected galaxies \citep{marin2010} and used in conjunction with dark-matter only simulations \citep{bagla2010, khandai2011, gong2011, guhasarkar2012} and with SPH simulations \citep{navarro2014}. 

It is important to be able to quantify and estimate the uncertainty in the various parameters that characterize the intensity fluctuation power spectrum, for the planning of current and future HI intensity mapping experiments. 
In this paper, we combine the presently available constraints on the neutral hydrogen gas mass density, $\Omega_{\rm HI}$ and bias parameter, $b_{\rm HI}$ to predict the subsequent uncertainty on the power spectrum of the 21-cm intensity fluctuations at various redshifts. The constraints are incorporated from galaxy surveys, HI intensity mapping experiments, the Damped Lyman Alpha system observations, theoretical prescriptions for assigning HI to dark matter haloes, and the results of numerical simulations. We find that it might be possible to improve upon the commonly used assumption of constant values of $\Omega_{\rm HI}$ and $b_{\rm HI}$ across redshifts by taking into consideration the fuller picture implied by the current constraints. {{We use a minimum variance interpolation scheme to obtain the uncertainties in $\Omega_{\rm HI}$ and $b_{\rm HI}$ across redshifts from 0 to $\sim 3.5$. We consider three different confidence scenarios for incorporating observational data and theoretical predictions.}} We discuss the resulting uncertainty in the HI power spectrum and the consequences for its measurement by current and future intensity mapping experiments. We also provide a tabular representation of the uncertainties in $\Omega_{\rm HI}$, $b_{\rm HI}$ and the power spectrum across redshifts, implied by the combination of the current constraints.

The paper is organized as follows. In Sec. \ref{sec:formalism}, we describe in brief the theoretical formalism leading to the 21-cm intensity fluctuation power spectrum and the ingredients that introduce sources of uncertainty. In Sec. \ref{sec:constraints}, we summarize the current constraints for the parameters in the power spectrum from the observational, theoretical and simulation results that are presently available. In the next section, we combine these constraints to obtain the uncertainty on the product $\Omega_{\rm HI} b_{\rm HI}$ which directly relates to the uncertainty in the power spectrum discussed in in Sec. \ref{sec:impact}. We summarize our findings and discuss future prospects in the final concluding section.

 \section{Formalism}
 \label{sec:formalism}
 
 \subsection{HI intensity mapping experiments}
 
 In the studies of 21-cm intensity mapping, the main observable is the three-dimensional power spectrum of the intensity fluctuation, $[\delta T_{\rm HI}(k,z)]^2$,  given by the expression \citep[e.g.,][]{battye2012}:
\begin{equation}
 [\delta T_{\rm HI}(k,z)]^2 = \bar{T}(z)^2 [b_{\rm HI}(k,z)]^2 \frac{k^3 P_{\rm cdm} (k,
z)}{2 \pi^2}
\label{3dpowspec}
\end{equation} 
where the mean brightness temperature is given by
\begin{eqnarray}
 \bar T (z) &=& \frac{3 h_{\rm Pl} c^3 A_{10}}{32 \pi k_B m_p^2 \nu_{21}^2}\frac{(1+z)^2}{H(z)} \Omega_{\rm HI} (z) \rho_{c,0} \nonumber \\
            &\simeq& 44 \ \mu {\rm{K}} \left(\frac{ \Omega_{\rm HI} (z) h }{2.45 \times10^{-4}}\right)\frac{(1+z)^2}{E(z)},
  \label{tbar}
\end{eqnarray}
where $E(z) = H(z) / H_0$, $b_{\rm HI}(k,z)$ is the HI bias, $\rho_{c,0}$ is the critical density at the present epoch ($z = 0$)  and $P_{\rm cdm} (k, z)$ is the
dark matter power spectrum, $A_{10}$ is the Einstein-A coefficient for the spontaneous emission between the lower (0) and upper (1) levels of hyperfine splitting, $\nu_{21}$ is the frequency corresponding to the 21-cm emission, and other symbols have their usual meanings. The above expression is calculated by assuming that the line profile, $d\nu$, is very narrow and absorption is neglected (which is a valid approximation if the spin temperature of the gas is far greater than the background CMB temperature). Also, it is assumed that the line width $d\nu/(1+z)$ is much smaller than the frequency interval of the observation.\footnote{We do not take into account peculiar velocity-related effects in the present study.}

As can be seen, the two key inputs to the power spectrum are the neutral hydrogen density parameter, $\Omega_{\rm HI}(z)$ and the bias parameter of HI, $b_{\rm HI}(k,z)$. These represent fundamental quantities in the observations of the HI intensity. In what follows, we will neglect the scale dependence of bias and treat it as a function of the redshift $z$ alone, i.e. $b_{\rm HI} (z)$. This is a valid approximation on large scales where we study the effects on the power spectrum.


\subsection{Halo model : Analytical calculation of $\lowercase{b}_{\rm HI}$ and $\Omega_{\rm HI}$}
\label{sec:analytical}
Here, we briefly outline the analytical formulation using the halo model for the distribution of dark matter haloes, which we use to compute the two quantities $\Omega_{\rm HI}$ and $b_{\rm HI}$. The Sheth-Tormen prescription \citep{sheth2002} for the halo mass function, $dn(M,z)/dz$,  is used for modelling the distribution of dark matter haloes. The dark matter halo bias $b(M)$ is then given following  \citet{scoccimarro2001}.

Given a prescription for populating the halos with HI, i.e. $M_{\rm HI}(M)$, defined as the mass of HI contained in a halo of mass $M$, we can compute the comoving neutral hydrogen density, $\rho_{\rm HI}(z)$, as:
\begin{equation}
\rho_{\rm HI}(z)= \int_{M_{\rm min}}^{\infty} dM \frac{dn(M,z)}{dM} M_{\rm HI}(M) \, ,
\end{equation}
and the bias parameter of neutral hydrogen, $b_{\rm HI} (z)$ as:
\begin{equation}
b_{\rm HI} (z) = \frac{1}{\rho_{\rm HI}(z)}\int_{M_{\rm min}}^{\infty} dM
\frac{dn(M,z)}{dM} b(M) M_{\rm HI}(M).
\label{biasHI}
\end{equation}
 We consider only the linear bias in this paper. 
 
Finally, the neutral hydrogen fraction is computed as (following common convention):
\begin{equation}
 \Omega_{\rm HI} (z) = \frac{\rho_{\rm HI}(z)}{\rho_{c,0}}
 \label{omegaHIanalyt}
\end{equation} 
where $\rho_{c,0} \equiv 3 H_0^2/8 \pi G$ is the critical density at redshift 0.

In the above analytical calculation, we see that the key input is $M_{\rm HI}(M)$, the prescription for
assigning HI to the dark matter haloes. This is done in several ways in the
literature and the various resulting prescriptions are discussed below and compiled in the lower section of Table
\ref{table:constraints}. These prescriptions have been found to be a good match
to observational results. We also consider the distribution of HI in haloes
resulting from smoothed-particle hydrodynamical simulations \citep{dave2013}. 

Given the values of $\Omega_{\rm HI}$ and $b_{\rm HI}$, the HI power spectrum may be computed following \eq{3dpowspec}. We do this using the linear matter power spectrum and the growth function obtained by solving its differential equation \citep{wang1998, linder2003, komatsu2009}. The cosmological
parameters assumed here are $\Omega_{\Lambda} = 0.723$, $h = 0.702$, $\Omega_{\rm m} = 0.277$,
$Y_p = 0.24$, $n_s = 0.962$, $\Omega_{\rm b} h^2 = 0.023$ which are in roughly good agreement with most available observations including the latest Planck results \citep{planck}. The primordial power spectrum corresponds to the normalization
$\sigma_8 = 0.815$. The matter transfer function is obtained from the fitting formula of \citet{eisenstein1998} including
the effect of baryonic acoustic oscillations. 

\subsection{Damped Lyman Alpha systems}

In studies measuring the neutral hydrogen fraction using Damped Lyman Alpha absorption systems (DLAs), the key observables are the sum $\Sigma N_{\rm HI}$ of the measurements of the column density of HI over a redshift interval having an absorption path length $\Delta X$, defined following \citet{lanzetta1991}. From this, the gas density parameter $\Omega_g^{\rm DLA}$ is evaluated as:
\begin{equation}
 \Omega_g^{\rm DLA} = \frac{\mu m_H H_0}{c \rho_{c,0}} \frac{\Sigma N_{\rm HI}}{\Delta X}
\end{equation} 
which is the discrete$-N$ limit of the exact integral expression:
\begin{equation}
 \Omega_g^{\rm DLA} = \frac{\mu m_H H_0}{c \rho_{c,0}} \int_{N_{\rm HI, min}}^{\infty} N_{\rm HI} f_{\rm HI} (N_{\rm HI}, X) dN_{\rm HI} dX \, ,
\end{equation} 
where the lower limit of the integral is set by the column density threshold for DLAs, i.e. $N_{\rm HI, min} = 10^{20.3}$ cm$^{-2}$. In case the sub-DLAs too are accounted for while calculating the gas density parameter, the same limit is usually taken to be $10^{19}$ cm$^{-2}$ \citep{zafar2013}. The low column-density systems, e.g., the Lyman-$\alpha$ forest make negligible contribution to the total gas density.
In the above expression, $\mu$ is the mean molecular weight, $m_H$ is the mass of the hydrogen atom, and $\rho_{c,0}$ is the critical mass density of the universe at redshift 0.
Also, $f_{\rm HI} (N, X)$ is the distribution function of the DLAs, defined through:
\begin{equation}
 d^2 \mathcal{N} = f_{\rm HI} (N_{\rm HI}, X) dN dX
\end{equation} 
with $\mathcal{N}$ being the incidence rate of DLAs in the absorption interval $dX$ and the column density range $dN_{\rm HI}$. Once  $\Omega_g^{\rm DLA}$ is known at several redshifts, it is possible to compute the hydrogen neutral gas mass density parameter $\Omega_{\rm HI}^{\rm DLA}$ for an assumed helium fraction by mass. This represents the neutral hydrogen fraction from DLAs alone. The bias parameter $b_{\rm DLA}$ for DLAs may be obtained from cross-correlation studies \citep{fontribera2012} with the Lyman-$\alpha$ forest.

Thus, the two parameters $\Omega_{\rm HI}$ and $b_{\rm DLA}$ may be estimated from DLA observations. However, as we see above, the techniques for the analysis of the DLA observations are different from those used in the galaxy surveys and HI intensity mapping experiments, both in terms of the fundamental quantities and the methods of calculation of $\Omega_{\rm HI}$ and $b_{\rm DLA}$. It was recently shown, using a
combination of SPH simulations and analytical prescriptions for assigning HI to haloes, that it is possible to model the 21 cm signal which is consistent with observed measurements of $\Omega_{\rm HI}$ and $b_{\rm DLA}$ \citep{navarro2014}.

 \section{Current constraints}
 \label{sec:constraints}
  Table
\ref{table:constraints} lists the presently available observational and
theoretical constraints on the various quantities related to the
computation of the HI three-dimensional power spectrum at different redshifts. The details of the various constraints are briefly described in the following.

\begin{table*}
\begin{center}
 \hspace{0in}  \begin{tabular}{llll}
    \hline
     Technique   & Constraints  & Mean redshift (Redshift range) & Reference\\
\hline\hline
      Observational              & &  & \\ \hline
\noalign{\smallskip}            
      Galaxy surveys                           &  & & \\ \hline
      ALFALFA 21-cm emission & $\Omega_{\rm HI}^{*} = 3.0 \pm 0.2$  & 0.026 & \citet{martin10}
 \\
\noalign{\smallskip} 
      HIPASS 21-cm emission & $\Omega_{\rm HI} = 2.6 \pm 0.3$ & 0.015  & \citet{zwaan05}\\
\noalign{\smallskip} 
      HIPASS, Parkes; HI stacking & $\Omega_{\rm HI} = 2.82^{+0.30}_{-0.59}$ & 0.028 (0 - 0.04) & \\
      & $\Omega_{\rm HI} = 3.19^{+0.43}_{-0.59}$ &  0.096 (0.04 - 0.13) &
 \citet{delhaize13} \\
\noalign{\smallskip} 
      AUDS (preliminary) &   $\Omega_{\rm HI} =  3.4 \pm 1.1$ & 0.125 (0.07 - 0.15) & 
 \citet{freudling11} \\
\noalign{\smallskip} 
      GMRT 21-cm emission stacking & $\Omega_{\rm HI} = 4.9 \pm 2.2$ & 0.24  & 
\citet{lah07} \\
\noalign{\smallskip}
      HI distribution maps from M31, M33 & & \\
      and LMC & $\Omega_{\rm HI} = 3.83 \pm 0.64$ & 0.0 & \citet{braun2012} \\
\noalign{\smallskip} 
      ALFALFA $\alpha$.40 sample, Millennium & $b_{\rm HI} = 0.7 \pm 0.1$ & \\
      simulation & (large scales) & $\sim 0$ & \citet{martin12} \\
\noalign{\smallskip}
\noalign{\smallskip} \hline
DLA observations    & & & \\ \hline
      DLA measurements  & $\Omega_{\rm HI} = 5.2 \pm 1.9$    & 0.609 (0.11 - 0.90)  &  \\
      from HST and SDSS & $\Omega_{\rm HI} = 5.1 \pm 1.5$    & 1.219 (0.90 - 1.65) & \citet{rao06}\\
      & $\Omega_{\rm HI} = 4.29^{+0.24}_{-0.23}$ & (2.2 - 5.5) &  \citet{prochaska09} \\
       & $\Omega_{\rm HI}(z)$ & (2.0 - 5.19)  & \citet{noterdaeme09, noterdaeme12} \\ 
\noalign{\smallskip} 
      Cross-correlation of DLA and Ly-$\alpha$ forest & & \\
      observations & $b_{\rm DLA} = 2.17 \pm 0.2 $ &  $\sim 2.3$ &\citet{fontribera2012} \\
\noalign{\smallskip}
     Observations of DLAs with HST/COS & $\Omega_{\rm HI} = 9.8^{+9.1}_{-4.9}$ & $< 0.35$ & \citet{meiring2011} \\
\noalign{\smallskip} 
     DLAs and sub-DLAs with VLT/UVES & $\Omega_{\rm HI}(z)$ & 1.5 - 5.0 & \citet{zafar2013} \\
\noalign{\smallskip} \hline
HI intensity mapping & & & \\ \hline
     WSRT HI 21-cm emission,  & $\Omega_{\rm HI} = 2.31 \pm 0.4$ & 0.1 & \\
     $z = 0.1$ \& 0.2 &  $\Omega_{\rm HI} = 2.38 \pm 0.6$ &   0.2 & \citet{rhee13}\\
\noalign{\smallskip} 
      Cross-correlation of DEEP2 galaxy-HI & & \\
      fields at $z = 0.8$ & $\Omega_{\rm HI} b_{\rm HI} r^{\dagger} = (5.5 \pm 1.5)h$ & 0.8 & \citet{chang10} \\
 \noalign{\smallskip} 
      21 cm intensity fluctuation
      cross-correlation with WiggleZ & & & \\
      survey &         $\Omega_{\rm HI} b_{\rm HI} r = (4.3 \pm 1.1)h$        & 0.8 & \citet{masui13} \\      
\noalign{\smallskip} 
      Auto-power spectrum of HI  & & & \\
      intensity field combined with &  $\Omega_{\rm HI} b_{\rm HI} = 6.2^{+2.3}_{-1.5}h$ &  &\\
      cross-correlation with WiggleZ &  & & \\
      survey &  & 0.8 &  \citet{switzer13}\\ 
 \hline
\noalign{\bigskip} 
\noalign{\smallskip} 
      Theory/Simulation  & &  &  \\ \hline
\noalign{\smallskip} 
      SPH simulation using GADGET-2 &  $M_{\rm HI}/M_{\rm halo}$, \\
      & $M_{*}/M_{\rm halo}, \Omega_{\rm HI} (z)$ &  $\sim
0$  & \citet{dave2013} \\ 
\noalign{\smallskip}
      Hydrodynamical simulation using GADGET-2/OWLs & $\Omega_{\rm HI} = (1.4 \pm 0.18) h$ & 0 & \\
      &  $\Omega_{\rm HI} = (2.5 \pm 0.14) h$ & 1 & \citet{duffy2012} \\
      &  $\Omega_{\rm HI} = (3.8 \pm 0.08) h$ & 2  & \\
\noalign{\smallskip} 
      N-body simulation, HI prescription & $b_{\rm HI}(k,z)$ & $ \sim 1.5 - 4$ & \citet{guhasarkar2012},\\
& & $\sim 1.3, 3.4$ and $5.1$ & \citet{bagla2010}  \\
\noalign{\smallskip}
      N-body simulation, HI prescription & $\Omega_{\rm HI} = (11.2 \pm 3.0) h$ & &\\
      combined with \citet{chang10} & $b_{\rm HI} = 0.55 - 0.65$  &   $ \sim 0.8$ & \citet{khandai2011}  \\
\noalign{\smallskip}
       Non-linear fit to the  & & & \\
       simulations of \citet{obreschkow2009} & $M_{\rm HI}/M_{\rm halo},M_{*}/M_{\rm halo}$ &  $ 1, 2, 3$ & \citet{gong2011}\\
\noalign{\smallskip} 
      HI prescription incorporating & & \\
      observational constraints & $b_{\rm HI}(z)$ & 0.0 -
3.0 & \citet{marin2010} \\
\noalign{\smallskip}
\hline
$*$ The units of $\Omega_{\rm HI}$ are $h^{-1} \times 10^{-4}$. \\
$\dagger$ Here, $r$ denotes the stochasticity. & & & 
  \end{tabular} 
  \end{center}
   \caption{Presently available constraints on the various
quantities required for calculation of the HI 3D power spectrum at different
redshifts. Constraints are broadly grouped into observational and
theoretical/simulation. Observational constraints include those from galaxy surveys, DLA observations and HI intensity mapping experiments. The columns list the
technique, parameter(s) constrained, the mean
redshift/redshift range, where available, and the reference in the literature for each. }

\label{table:constraints}
\end{table*}
 \subsection{Observational}
The top half of Table \ref{table:constraints} summarizes the current observational constraints, which are briefly described below:
\begin{itemize}
\item Galaxy surveys:

The Arecibo Fast Legacy ALFA Survey (ALFALFA) surveys 21-cm emission lines 
from a region of 7000 deg$^2$, producing deep maps of the HI distribution in the 
local universe out to redshift $z \sim 0.06$. \citet{martin10} use a sample of 10119 HI-selected galaxies from the $\alpha.40$ survey to calculate the HI mass function and find the cosmic neutral HI gas density $\Omega_{\rm HI}$ at $z = 0$.
In \citet{martin12},  the correlation function of HI-selected galaxies in the local universe measured by the $\alpha.40$ survey, together with the correlation function of dark matter haloes as obtained from  the Millennium simulation \citep{millenium2005}, is used to estimate the bias parameter $b_{\rm HI}$ in the local universe. 

\citet{zwaan05} present results of the measurement
of the HI mass function from the 21 cm emission-line detections of the HIPASS
catalogue whose survey measured the HI Mass Function (HIMF) and the neutral hydrogen fraction from 4315 detections of 21-cm line emission in a sample of  HI-selected galaxies in the local universe. This measurement is further used to estimate the neutral hydrogen mass density $\Omega_{\rm HI}$ in the local universe.

\citet{lah07} present 21-cm HI emission-line measurements using co-added observations from the Giant Meterwave Radio Telescope (GMRT) at redshift $z = 0.24$. This allows the estimation of the cosmic neutral gas density which can be converted into an estimate for $\Omega_{\rm HI}$ at this redshift.

\citet{braun2012} use high-resolution maps of the HI distribution in M31, M33 and the Large Magellanic Cloud (LMC), with a correction to the column density based on opacity, to constrain the neutral hydrogen gas mass density at $z = 0$. 

\citet{delhaize13} use the HIPASS and the Parkes observations of the SGP field to place constraints on $\Omega_{\rm HI}$ at two redshift intervals, (0 - 0.04) and (0.04 - 0.13). 

\citet{rhee13} use the HI 21-cm emission line measurements of field galaxies with the Westerbork Synthesis Radio telescope (WSRT) at redshifts of 0.1 (59 galaxies) and 0.2 (96 galaxies) to measure the neutral hydrogen gas density at these redshifts.

\citet{freudling11} use a set of precursor observations of 18  21-cm emission lines at redshifts between redshifts 0.07 and 0.15 from the ALFA Ultra Deep Survey (AUDS) to derive the HI density $\rho_{\rm HI}$ at the median redshift 0.125. 

\item Damped Lyman Alpha systems (DLAs) observations:

\citet{rao06} use the HST and SDSS measurements of Damped Lyman Alpha systems (DLAs) at redshift intervals 0.11 - 0.90 (median redshift 0.609) and 0.90 - 1.65 (median redshift 1.219) to constrain the value of $\Omega_{\rm HI}$ at these epochs. 

 \citet{prochaska09} use a sample of 738 DLAs from SDSS-DR5, at redshifts 2.2 - 5.5, in six redshift bins to constrain the neutral hydrogen gas mass density. 
\citet{noterdaeme09} use  937 DLA systems from SDSS-II DR7 in four redshift bins from 2.15 - 5.2,  \citet{noterdaeme12} measure $\Omega_{\rm HI}^{\rm DLA}(z)$ using a sample of 6839 DLA systems from the Baryonic Oscillation Spectroscopic Survey (BOSS) which is part of the SDSS DR9, in five redshift bins between redshifts 2.0 and 3.5.

\citet{meiring2011} present the first observations from HST/COS of three DLAs and four sub-DLAs to measure the neutral gas density at $z < 0.35$.

\citet{fontribera2012} use the cross-correlation of DLAs and the Lyman-$\alpha$ forest to constrain the bias parameter of DLAs, $b_{\rm DLA}$ at redshift $z \sim 2.3$.

\citet{zafar2013} use the observations of DLAs and sub-DLAs from 122 quasar spectra using the European Southern Observatory (ESO) Very Large Telescope/Ultraviolet Visual Echelle Spectrograph (VLT/UVES), in conjunction with other sub-DLA samples from the literature, to place constraints on the neutral hydrogen gas mass density at $1.5 < z < 5$. One of the crucial differences between this work and others, e.g., \citet{noterdaeme12}, is that it accounts for sub-DLAs while calculating the total gas mass.

\item HI intensity mapping experiments:

\citet{chang10} used the Green Bank Telescope (GBT) to record radio spectra across two of the DEEP2 optical redshift survey fields and present a three-dimensional 21-cm intensity field at redshifts 0.53 - 1.12. The cross-correlation technique is used to infer the value of $\Omega_{\rm HI} b_{\rm HI} r$ (where $r$ is the stochasticity) at redshift $z = 0.8$.

In \citet{masui13}, the cross-correlation of the 21 cm intensity fluctuation with the WiggleZ survey is used to constrain $\Omega_{\rm HI} b_{\rm HI} r$. 

In \citet{switzer13}, the auto-power spectrum of the 21 cm intensity fluctuations is combined with the above cross-power treatment to constrain the product $\Omega_{\rm HI} b_{\rm HI}$ at $z \sim 0.8$.

\end{itemize}

 \subsection{Theoretical}
 The theoretical constraints arise from various prescriptions for assigning HI to dark matter haloes. These prescriptions, for different redshifts, are briefly summarized below.
 
 \begin{itemize}
  \item Redshift $\sim 0$: In \citet{dave2013}, Fig. 10 is plotted
$M_{\rm HI}(M)$ at $z = 0$ from their smoothed-particle
hydrodynamical simulation. We interpolate the values of $M_{\rm HI}(M)$ to obtain a smooth curve.

\item Redshift $\sim 0$: The prescription in \citet{marin2010} uses a fit
to the observations of \citet{zwaan05} and gives $M_{\rm HI}$ as a
function of $M$ at redshift $z \sim 0$.

\begin{figure}
 \begin{center}  
\includegraphics[scale = 0.33, angle = 90]{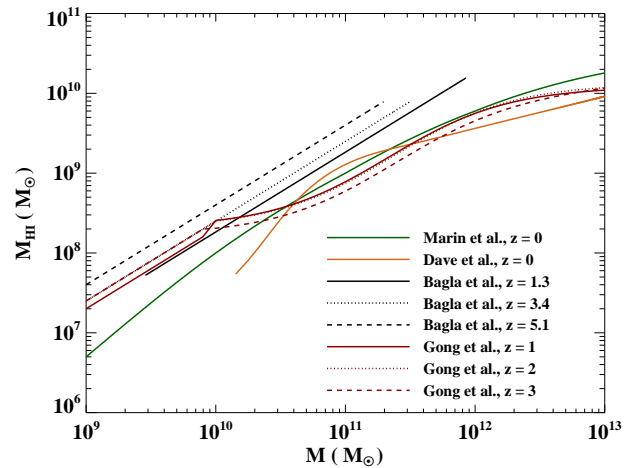} 
 \end{center}
\caption{Prescriptions from the literature for assigning HI to dark matter haloes. Results from \citet{dave2013}, \citet{marin2010} at redshift $\sim 0$, \citet{bagla2010} at redshifts 1.3, 3.4 and 5.1 and \citet{gong2011} at redshifts 1, 2 and 3 give $M_{\rm HI}$ as a function of the halo mass $M$.}
\label{fig:bagla}
\end{figure}

\begin{figure*}
 \begin{center}  
\hskip-0.2in \includegraphics[scale = 0.53, angle = 90]{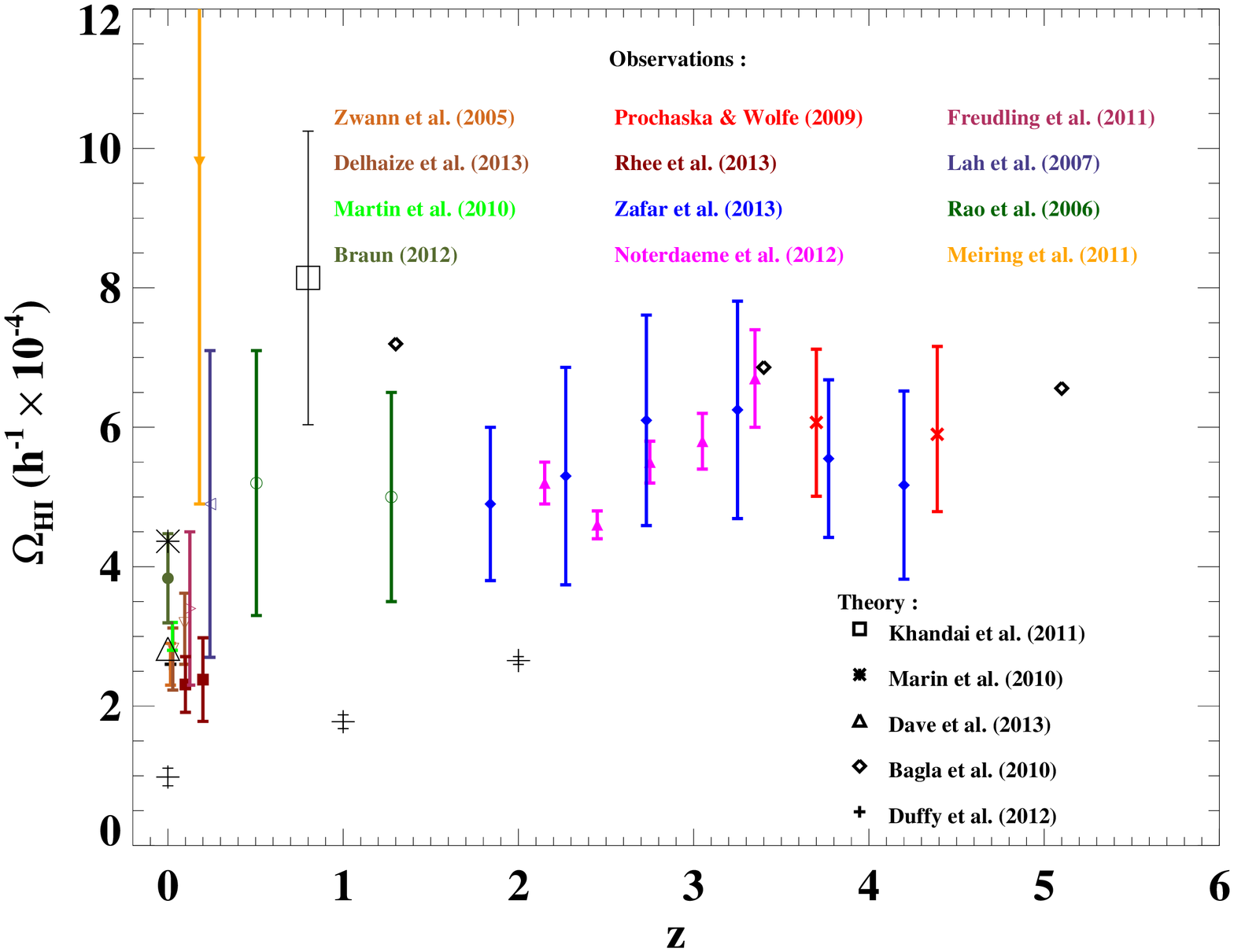} 
 \end{center}
\caption{Compiled $\Omega_{\rm HI}$ values in units
of $h^{-1} \times 10^{-4}$ from the literature : the observations of \citet{zwaan05} (chocolate brown solid line), \citet{braun2012} (olive filled circle), \citet{delhaize13} (brown open downward triangles), \citet{martin10} (green dot), \citet{freudling11} (maroon right triangle),
\citet{lah07} (purple left triangle), \citet{rao06} (dark green open circles), \citet{prochaska09} (red crosses), \citet{rhee13} (dark red filled squares), \citet{meiring2011} (orange filled downward triangle), \citet{noterdaeme12} (magenta filled triangles), \citet{zafar2013} (blue filled diamonds), and the theoretical/simulation
prescription predictions of \citet{khandai2011}, \citet{marin2010, dave2013}, \citet{bagla2010} and \citet{duffy2012}. The observational points are plotted in color and the theoretical ones in black.}
\label{fig:omegaHI}
\end{figure*}

\item Redshifts $z > 0$ : The prescription given by \citet{bagla2010} assigns a constant ratio of HI mass to halo mass at each redshift,
denoted by $f_1$. The constant $f_1$ depends on the redshift under
consideration. For each of the three redshifts considered, $z = 1.5, 3.4$ and 5.1, the
value of $f_1$ is fixed by setting the neutral hydrogen density $\Omega_{\rm HI}$ to $10^{-3}$ in the simulations. The maximum and minimum masses of haloes containing HI gas are also redshift dependent. It is assumed
that haloes with masses corresponding virial velocities of less than 30 km/s and greater than 200
km/s are unable to host HI. \citet{guhasarkar2012} use the above prescription with the results of their N-body simulation to provide a cubic polynomial fit to the $b_{\rm HI}(k)$ at different redshifts.

A prescription for assigning HI to dark matter haloes at redshift $\sim 1$, for three different theoretical models has been presented in \citet{khandai2011}, for consistency with the observational constraints of $\Omega_{\rm HI}$ at $z \sim 0.8$ \citep{chang10}. This is used with an N-body simulation to predict, in conjunction with the results of \citet{chang10}, the neutral hydrogen density $\Omega_{\rm HI}$ and the bias factor $b_{\rm HI}$ at this redshift.

\citet{gong2011} provide non-linear fitting functions for assigning HI to dark matter haloes at redshifts $z \sim 1, 2$ and 3, based on the results of the simulations generated by \citet{obreschkow2009}.

\citet{duffy2012} use results of high-resolution cosmological hydrodynamical simulations with the GADGET-2/OWLS including the modelling of feedback from supernovae, AGNs and a self-shielding correction in moderate density regions, in order to predict $\Omega_{\rm HI}$ at $z = 0, 1$ and 2.

The above prescriptions, where analytical forms are available, are plotted in Figure \ref{fig:bagla}. These functions are subsequently used to generate the bias and neutral hydrogen densities at the corresponding redshifts, $b_{\rm HI}$ and $\Omega_{\rm HI}$ as described in Section \ref{sec:formalism}.

\end{itemize}

\section{Combined uncertainty on $\Omega_{\rm HI}$ and $\lowercase{b}_{\rm HI}$}
\label{sec:combine}
 In this section, we compile the current constraints to formulate the combined uncertainty on the quantities $\Omega_{\rm HI}$ and $b_{\rm HI}$.

Figure \ref{fig:omegaHI} shows the compiled set of values of the
neutral hydrogen density parameter, $\Omega_{\rm HI}$ from the observations and
theory in Table \ref{table:constraints}. The theoretical points are obtained by using \eq{omegaHIanalyt} of the formalism described in Sec. \ref{sec:analytical} using the $M_{\rm HI} (M)$ prescriptions described in Sec. \ref{sec:constraints}.  \footnote{We set $M_{\rm min} = 10^9 h^{-1} M_{\odot}$ and $M_{\rm max} = 10^{13} h^{-1} M_{\odot}$ in all the computations except for those corresponding to the prescription of \citet{bagla2010} where the explicit values of $M_{\rm min}$ and $M_{\rm max}$ are specified for each redshift.}
The observational points are shown in color and the theoretical points are plotted in black.

In Figure \ref{fig:bias} are plotted the analytical estimates for the bias,
$b_{\rm HI}$ obtained by using \eq{biasHI} of the analytical formulation described in Sec. \ref{sec:analytical}, together
with the available prescriptions at the corresponding redshifts. These include :
(a) the
theoretical/simulation prescriptions of \citet{bagla2010}, \citet{marin2010}, \citet{dave2013}, \citet{gong2011}, and the fitting formula of \citet{guhasarkar2012} and (b) the measurements of the bias at $z \sim 0$ by the ALFALFA survey \citep{martin12}, the combined constraints in \citet{switzer13} and \citet{rao06} providing an estimate of $b_{\rm HI}$ at $z \sim 0.8$,\footnote{The statistical uncertainties are added in quadrature.}, and the value of $b_{\rm HI}$ at $z \sim 0.8$ estimated by \citet{khandai2011} using the measurement of \citet{chang10}. The theoretical values are plotted in black and the measurements are plotted in color.

We now use the compilation of the available measurements to obtain estimates on the values of $\Omega_{\rm HI}$ and $b_{\rm HI}$ at intervening redshifts, as also estimates on the 1$\sigma$ error bars at the intervening points. To do so, we need error estimates on all the data points for $\Omega_{\rm HI}$ and $b_{\rm HI}$. We use the observational points and their error bars as the data points in the case of $\Omega_{\rm HI}$. {{The case of $b_{\rm HI}$ is more speculative since there are very few observational constraints. The present constraints on $b_{\rm HI}$ include:

(a) the two available observations: the ALFALFA result at $z = 0$ from \citet{martin12}, and the combination of the \citet{switzer13} and the \citet{rao06} measurement at $z = 0.8$ with the corresponding error bars.

(b) the 10 theoretical points at $z > 1$.

To obtain estimates on the uncertainties in $b_{\rm HI}$, we may consider the following three scenarios:

(a) Conservative: In this approach, we may limit the analysis to the observational uncertainties on $b_{\rm HI}$, and neglect the theoretical predictions. We, therefore, may use only the two available observations, with their error bars, to constrain the bias. 

(b) Optimistic: In this alternate approach, we may consider the opposite situation, i.e. that the value of the bias is given by a theoretical model for all redshifts, with zero error. This in turn avoids the association of uncertainties to the theoretical predictions.

The above two scenarios (a) and (b) are considered further in the Appendix. 

(c) Intermediate scenario: We consider this scenario for the remainder of the main text. To motivate the approach, we re-emphasize that the analysis for the bias is dominated by theoretical and modelling uncertainties and hence, to fully utilize the available constraints, one needs to quantify the uncertainties in the modelling at each redshift. If the scatter in individual simulations is considered as an estimate of the error, the error bars in most cases turn out to be negligibly low (corresponding effectively to case (b) above) and also do not reflect the range of physics input that may be used in other simulations at the same redshift. Hence, one possible method is to use the range of values of bias predicted by all the available theoretical models at a certain redshift as a measure of the range of physics uncertainties in the theoretical models. Here, we use the 10 theoretical points at $z > 1$, with a binned average to calculate the mean and $1\sigma$ deviation in four redshift bins each of width $\sim 0.6$ between redshifts $1 < z < 3.5$. This serves as an estimate of the error due to modelling uncertainties in the calculation of the bias factor. In this way, we obtain estimates on the mean and error bars on the bias factor at redshifts $1 < z < 3.5$. The values and error bars for $\Omega_{\rm HI}$ and $b_{\rm HI}$ thus obtained are plotted in Figure \ref{fig:snake}.

We note that the scenarios (b) and (c) contain contributions from the results of simulations. The choice of physics in the simulations and their possible biases, therefore, have an influence on the results obtained and their uncertainties. The validity of the results may be confirmed when further data becomes available at higher redshifts.

We use the algorithm for interpolation of irregularly spaced noisy data using the minimum variance estimator as described in \citet{rybicki1992}. This estimator is so constructed that both the error as well as the spacing between the noisy data points are taken into consideration. As an input to the algorithm, one requires an estimate of the typical (inverse) decorrelation of the sample, $w$, which we take to be $w = 2$ that corresponds to a decorrelation length of 0.5 (in redshift units). We also assume the value of the \textit{a priori} population standard deviation $p_{\rm sig} = 2$.\footnote{The values of $w$ and $p_{\rm sig}$ are usually well-defined in the case of time-series data. Increasing the value of $w$ decreases the error on the estimate and vice versa. Similarly, increasing $p_{\rm sig}$ increases the error on the estimate and vice versa. We choose the values $w = 2$ and $p_{\rm sig} = 2$, since for these values of the decorrelation length and population standard deviation, the results obtained are visually a good fit to the data points, including the error estimates.}.  We implement the algorithm with the help of the fast tridiagonal solution described in \citet{rybicki1994}.\footnote{http://www.lanl.gov/DLDSTP/fast/} We thus obtain an estimate of the mean value and 1$\sigma$ error bars on intervening points for both $\Omega_{\rm HI}$ and $b_{\rm HI}$. These are plotted as the solid and dotted lines  (``snakes'') of Figures \ref{fig:snake}.

\begin{figure}
 \begin{center}  
\hskip-0.2in \includegraphics[scale = 0.33, angle = 90]{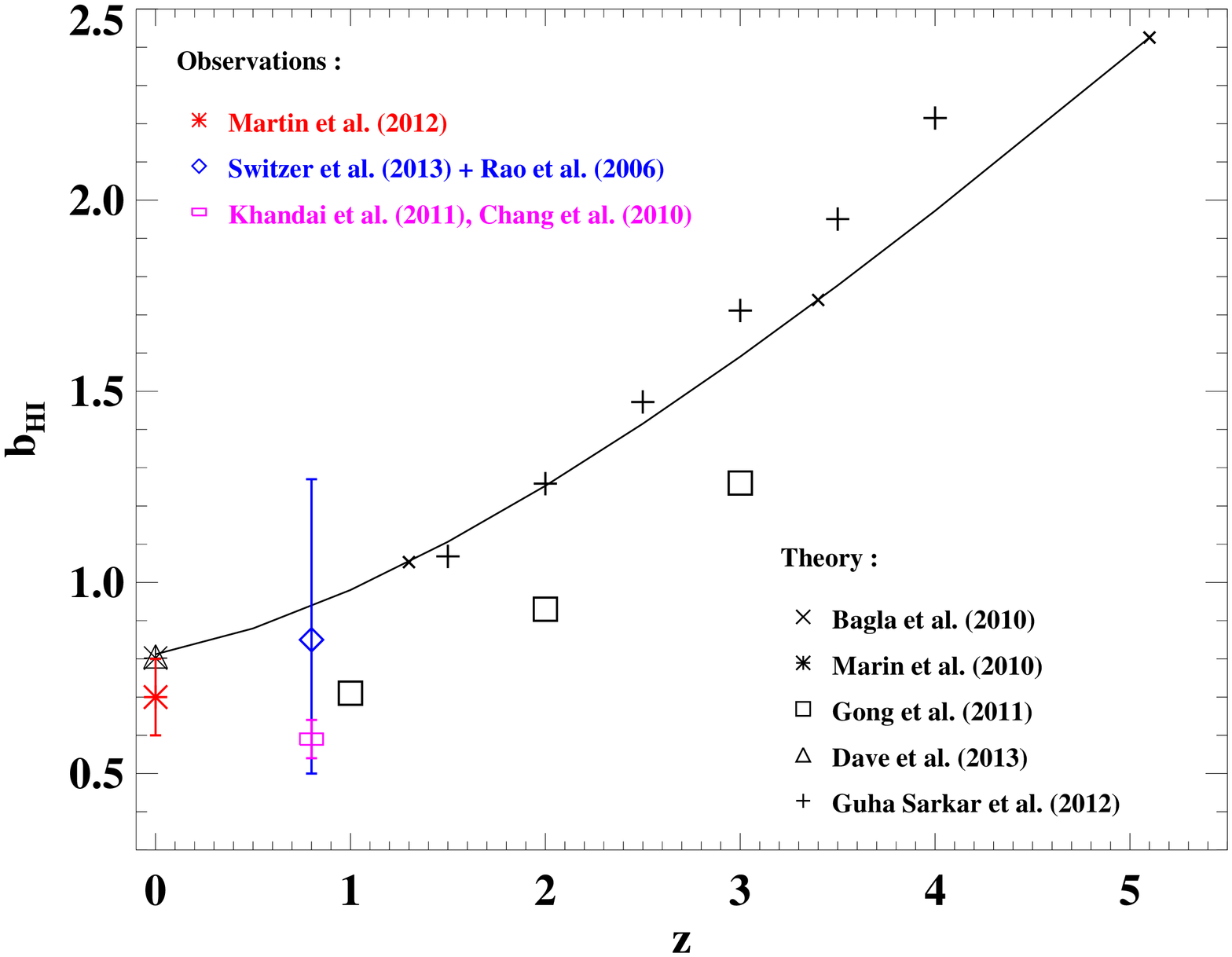} 
 \end{center}
\caption{The bias $b_{\rm HI}$ obtained from the
theoretical/simulation prescriptions of \citet{bagla2010}, \citet{marin2010}, \citet{dave2013}, \citet{gong2011}, and the fitting formula of \citet{guhasarkar2012} are shown in black. The solid black curve is the bias calculated using the theoretical prescription of \citet{bagla2010} at all redshifts under consideration, and is overplotted for reference.  The values of $b_{\rm HI}$ obtained by combining the observational results of \citet{switzer13} and \citet{rao06} at $z \sim 0.8$, the bias computed by \citet{khandai2011} at $z \sim 0.8$ using the observations of \citet{chang10}, and the bias value at $z \sim 0$ measured by \citet{martin12} for the ALFALFA sample of HI-selected galaxies are shown in color.}
\label{fig:bias}
\end{figure}

\begin{table*}
 \begin{center}
   \hspace{0in}  \begin{tabular}{lcllllll}
    \hline
    $z$   & $\Omega_{\rm HI} ^{\dagger}$ & $\Delta \Omega_{\rm HI}^{\dagger}$ & $b_{\rm HI}$ & $\Delta b_{\rm HI}$ & $\Omega_{\rm HI} b_{\rm HI}^{\dagger}$ & $\Delta (\Omega_{\rm HI}  b_{\rm HI})^{\dagger}$ & $\Delta (\Omega_{\rm HI}  b_{\rm HI})/(\Omega_{\rm HI} b_{\rm HI})$\\ 
   \hline 
   0.000         &   3.344         &   0.814         &   0.703         &   0.047         &   2.352         &   0.593         &   0.252        \\
   0.250         &   3.443         &   0.703         &   0.972         &   0.333         &   3.346         &   1.335         &   0.399        \\
   0.500         &   4.523         &   1.445         &   1.026         &   0.367         &   4.640         &   2.224         &   0.479        \\
   0.750         &   4.648         &   1.835         &   0.935         &   0.206         &   4.348         &   1.966         &   0.452        \\
   1.000         &   4.710         &   1.877         &   1.005         &   0.294         &   4.733         &   2.340         &   0.494        \\
   1.250         &   4.804         &   1.612         &   1.005         &   0.234         &   4.830         &   1.971         &   0.408        \\
   1.500         &   4.766         &   1.750         &   1.049         &   0.304         &   4.998         &   2.340         &   0.468        \\
   1.750         &   4.804         &   1.487         &   1.099         &   0.365         &   5.281         &   2.398         &   0.454        \\
   2.000         &   4.936         &   1.207         &   1.101         &   0.172         &   5.432         &   1.578         &   0.290        \\
   2.250         &   5.008         &   0.807         &   1.160         &   0.371         &   5.810         &   2.079         &   0.358        \\
   2.500         &   4.750         &   0.759         &   1.261         &   0.395         &   5.989         &   2.107         &   0.352        \\
   2.750         &   5.471         &   0.880         &   1.409         &   0.263         &   7.708         &   1.899         &   0.246        \\
   3.000         &   5.541         &   1.048         &   1.329         &   0.444         &   7.363         &   2.829         &   0.384        \\
   3.250         &   5.756         &   2.401         &   1.498         &   0.420         &   8.620         &   4.334         &   0.503        \\
   3.400         &   5.971         &   1.570         &   1.802         &   0.252         &  10.758         &   3.204         &   0.298        \\     
   \hline 
   $\dagger$ In units of $10^{-4} h^{-1}$. 
   \end{tabular}   
   \end{center}
 \caption{Combination of the fractional uncertainty on $b_{\rm HI}$, $\Omega_{\rm HI}$ due to the currently available constraints, and the predicted resulting uncertainty (both absolute and relative) on the product $\Omega_{\rm HI}  b_{\rm HI}$, which is the quantity of relevance for the calculation of the 3D temperature fluctuation power spectrum, $(\delta T_{\rm HI})^2$, at various redshifts. The range of interpolation is restricted upto redshift 3.4 due to the last bias point near $z \sim 3.4$. Note that $\Omega_{\rm HI}$ is in units of $10^{-4} h^{-1}$.}
\label{table:uncert}
 \end{table*}

The resulting values of the mean and errors in $\Omega_{\rm HI}$ and $b_{\rm HI}$ obtained by the interpolation of the data and the resulting estimate and uncertainty in the product $\Omega_{\rm HI} b_{\rm HI}$ are listed in Table \ref{table:uncert}.\footnote{The error estimates arise from a combination of (a) the magnitude of the errors on individual points as well as (b) the proximity to, and errors on, the nearby points. It can be seen that the errors at redshifts $z \sim 2.7$ are low, due to a number of nearby well-constrained points. In comparison, the errors near $z \sim 3.25$ are higher, due to the higher error bars on nearby points.}  These are also plotted in the curves of Figures \ref{fig:snake} and \ref{fig:snakeprod} along with the compiled data points and the measurement of $\Omega_{\rm HI} b_{\rm HI}$ at $z = 0.8$ \citep{switzer13}. {{These values are also fairly consistent with the uncertainties predicted by the conservative and optimistic scenarios over their ranges of applicability (see Appendix).
}}

\begin{figure}
 \begin{center}  
\includegraphics[scale = 0.32, angle = -90]{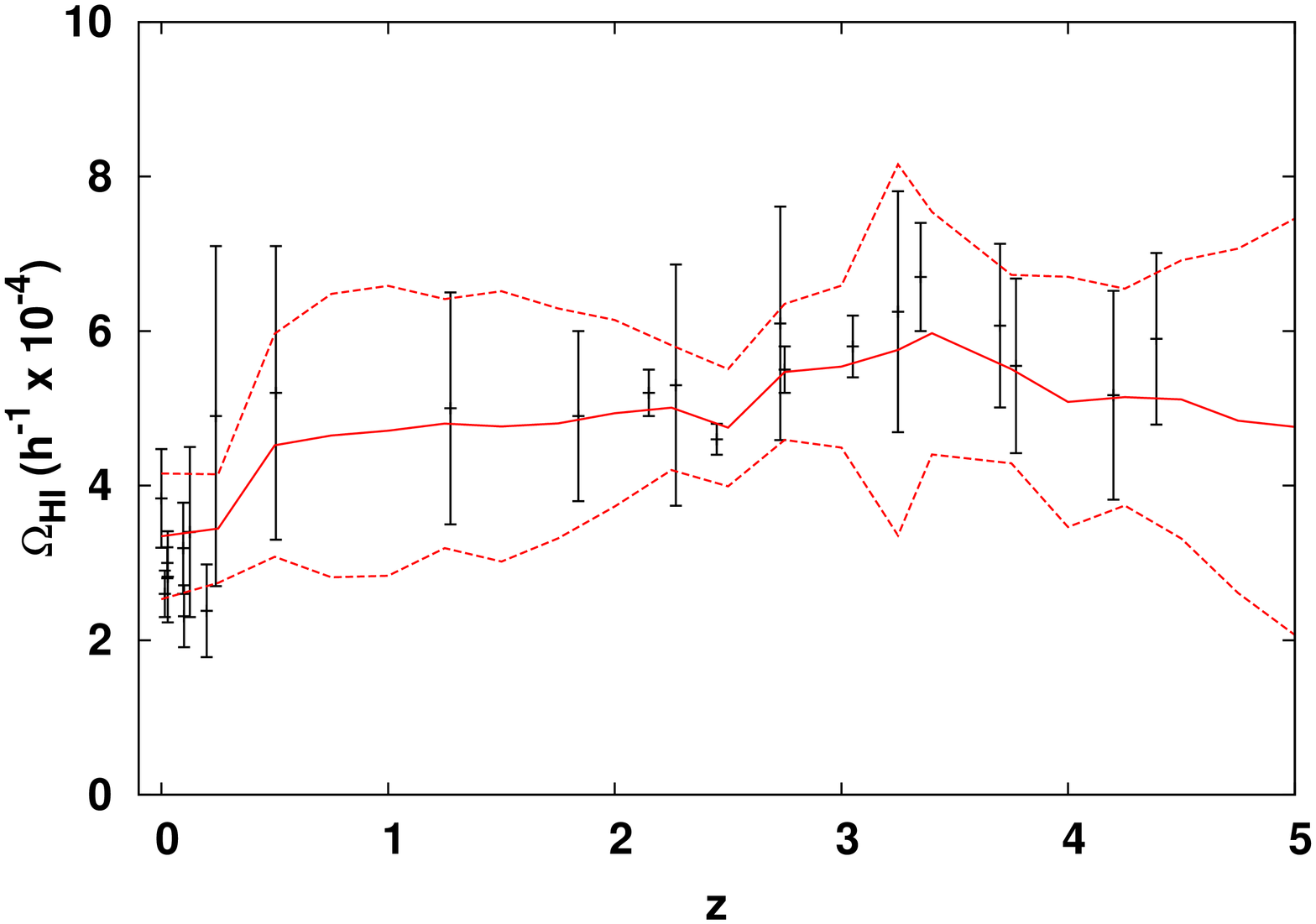} \includegraphics[scale = 0.32, angle = -90]{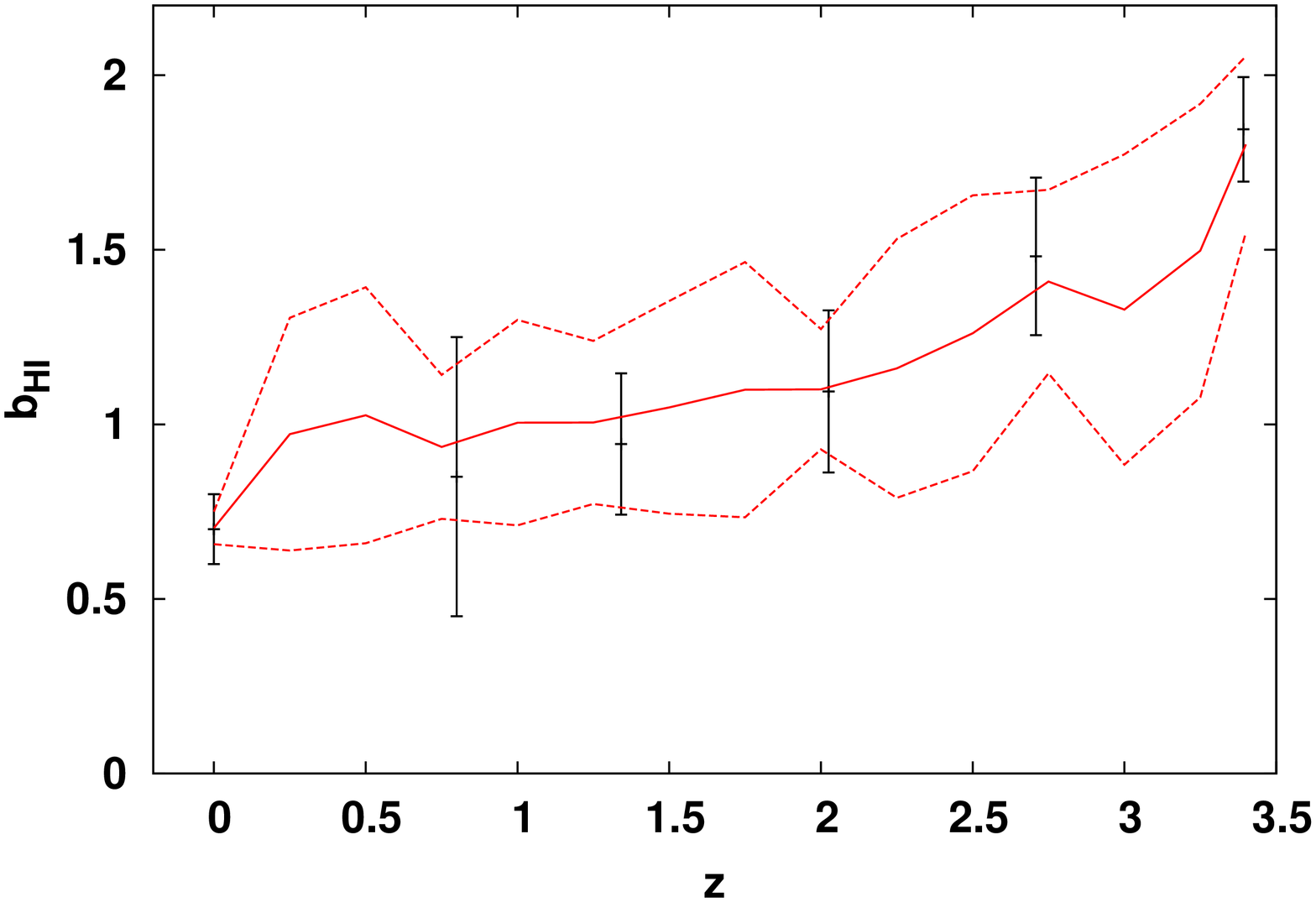}
 \end{center}
\caption{Compiled $\Omega_{\rm HI}$ values (in units
of $h^{-1} \times 10^{-4}$) and $b_{\rm HI}$  from the literature, with the minimum variance unbiased estimator (solid line) of \citet{rybicki1992} overplotted along with its 1$\sigma$ error in each case (dotted lines). In the case of $b_{\rm HI}$, the errors reflect the theoretical and modelling uncertainties and hence are more speculative. The bias $b_{\rm HI}$ is not as accurately constrained as $\Omega_{\rm HI}$ from observations, however the errors at present are dominated by the range of theoretical predictions for $b_{\rm HI}$ at different redshifts.}
\label{fig:snake}
\end{figure}

\begin{figure}
 \begin{center}  
\includegraphics[scale = 0.32, angle = -90]{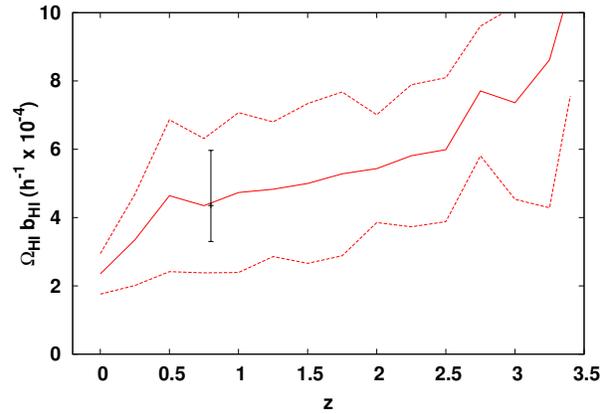}
 \end{center}
\caption{Compiled set of values of $\Omega_{\rm HI} b_{\rm HI}$ (in units
of $h^{-1} \times 10^{-4}$), calculated using the estimates for the mean and $1 \sigma$ standard deviations in $\Omega_{\rm HI}$ and $b_{\rm HI}$. The error estimate is obtained by propagating the errors in $\Omega_{\rm HI}$ and $b_{\rm HI}$. The measurement \citep{switzer13} at $z = 0.8$ is also overplotted for reference.}
\label{fig:snakeprod}
\end{figure}

 \section{Impact on the HI power spectrum}
 \label{sec:impact}
 
 As can be seen from Eqs. (\ref{3dpowspec}) and (\ref{tbar}), the quantity $\Omega_{\rm HI} b_{\rm HI}$ directly appears in the expression for the HI temperature fluctuation power spectrum. 
 Therefore, the HI temperature fluctuation and its power spectrum will be uncertain by different amounts depending upon the level of variation of $\Omega_{\rm HI}$ and $b_{\rm HI}$ allowed by observational and theoretical constraints. For example, at redshifts near 1, the temperature fluctuation varies by about 50\% due to the variation in the product $\Omega_{\rm HI} b_{\rm HI}$ alone. However, near redshifts 2 - 2.75, it is more constrained and varies only by about 25 - 35\% due to the larger number of tighter constraints on $\Omega_{\rm HI}$ at these redshifts. The power spectrum, $(\delta T_{\rm HI})^2$ has uncertainties of about twice this amount.
 Due to the very small number of data points above redshift 3.5, it is difficult to obtain constraints on $\delta T_{\rm HI}$ beyond this redshift with the presently available measurements. The  $\delta T_{\rm HI}$ and its resulting uncertainty are plotted for redshifts 0, 1, 2 and 3 in Figures \ref{fig:powspec}. 

\begin{figure}
  \begin{center}
   \includegraphics[scale = 0.32, angle = -90]{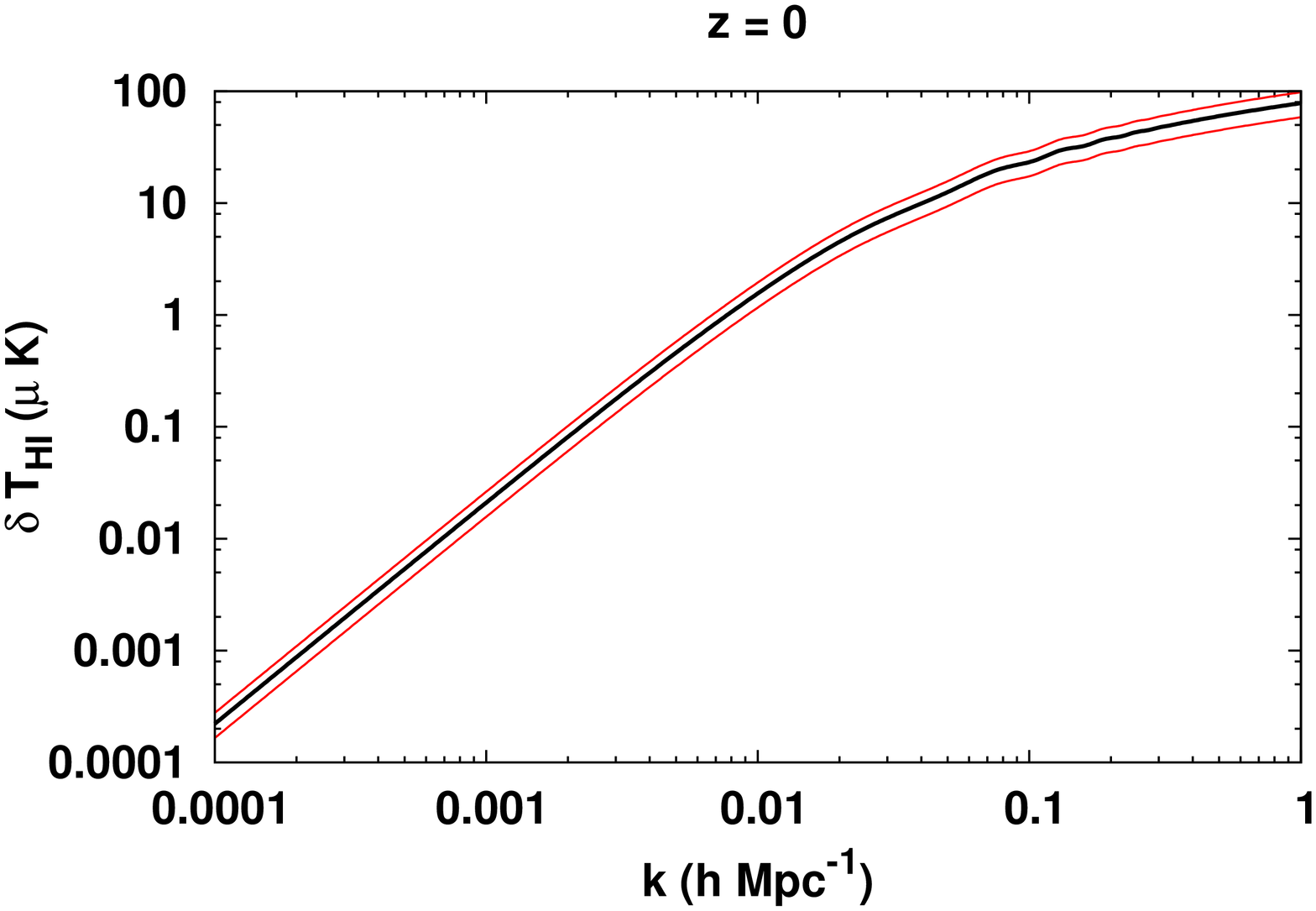} \includegraphics[scale = 0.32, angle = -90]{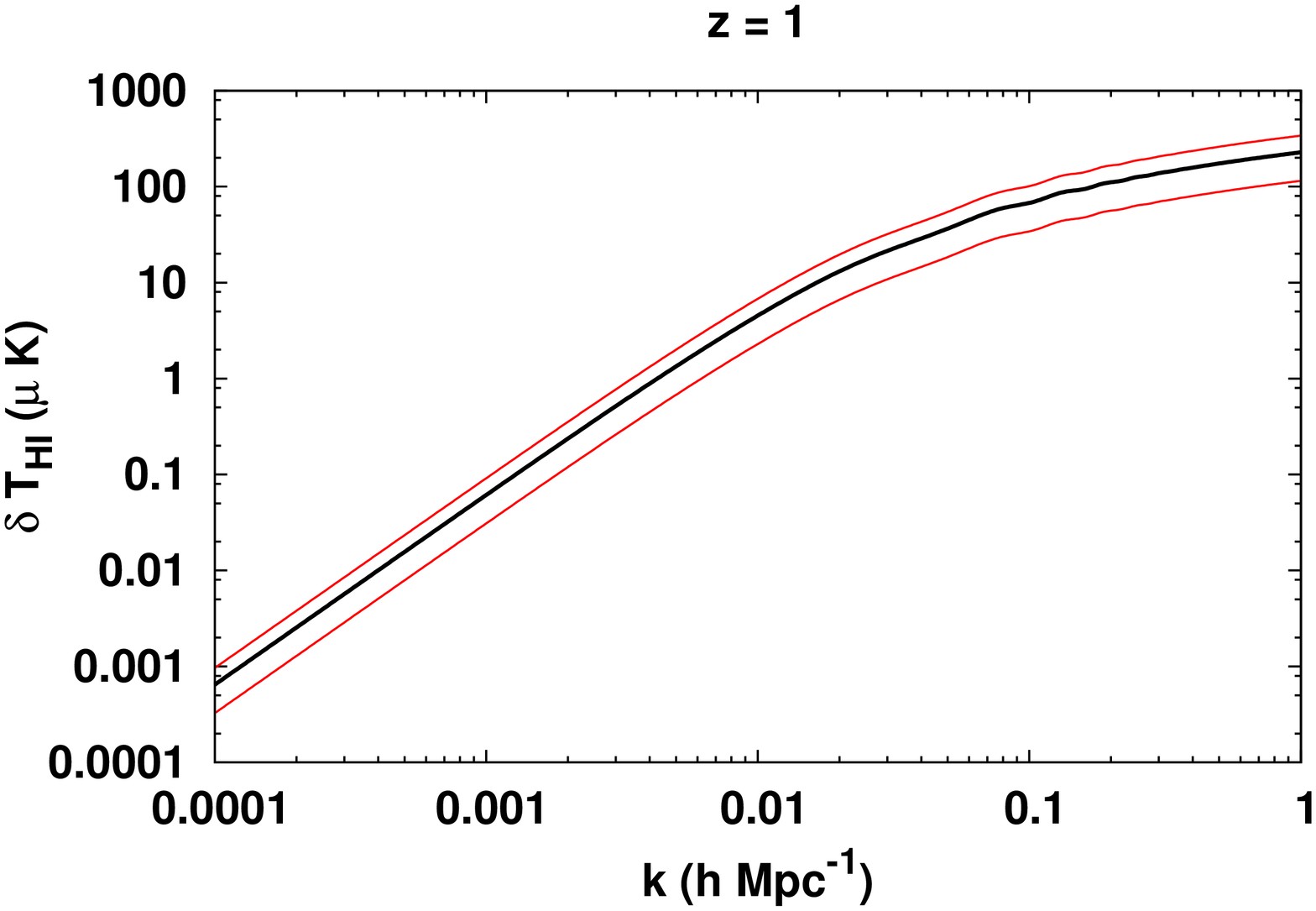} \includegraphics[scale = 0.32, angle = -90]{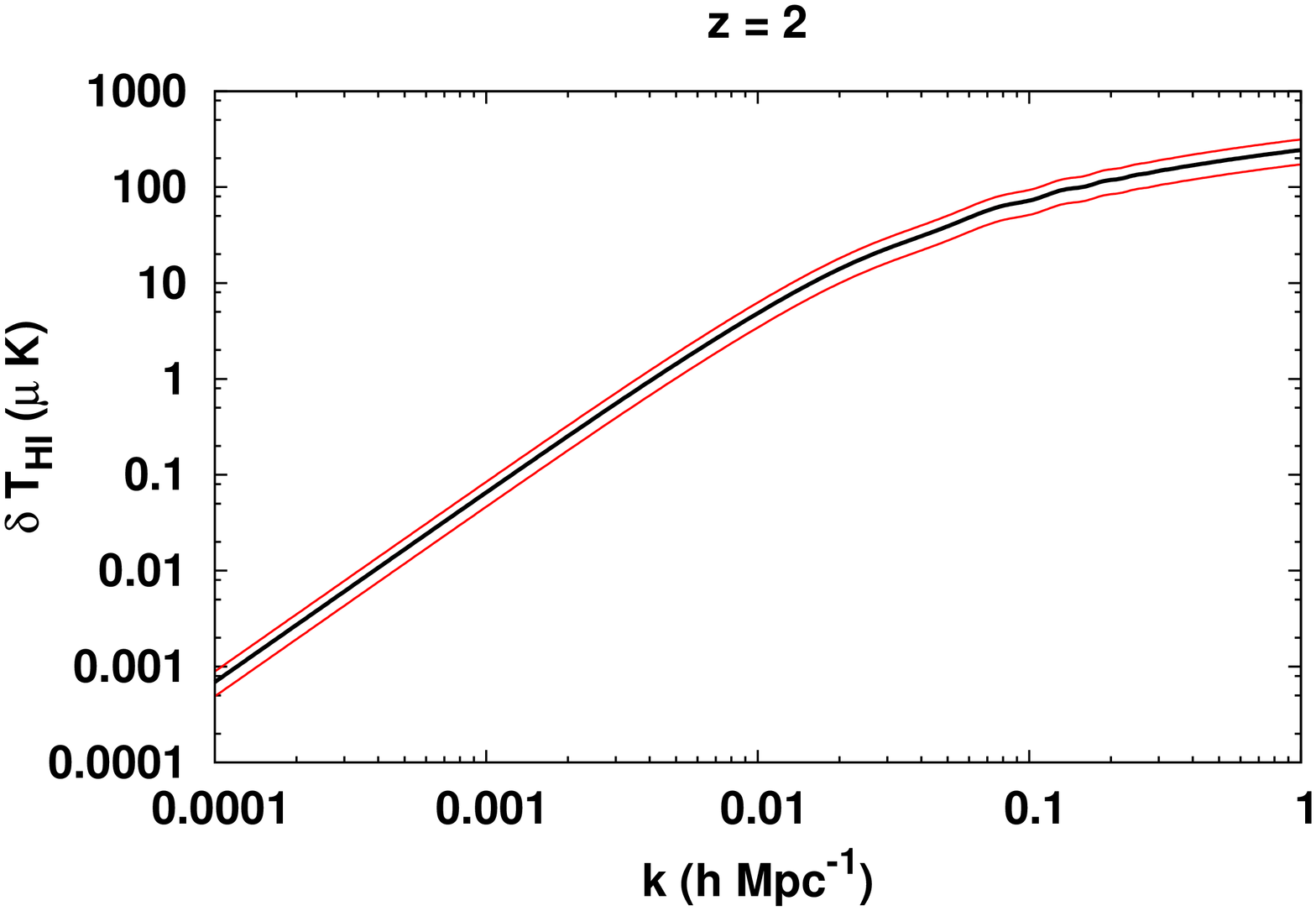} \includegraphics[scale = 0.32, angle = -90]{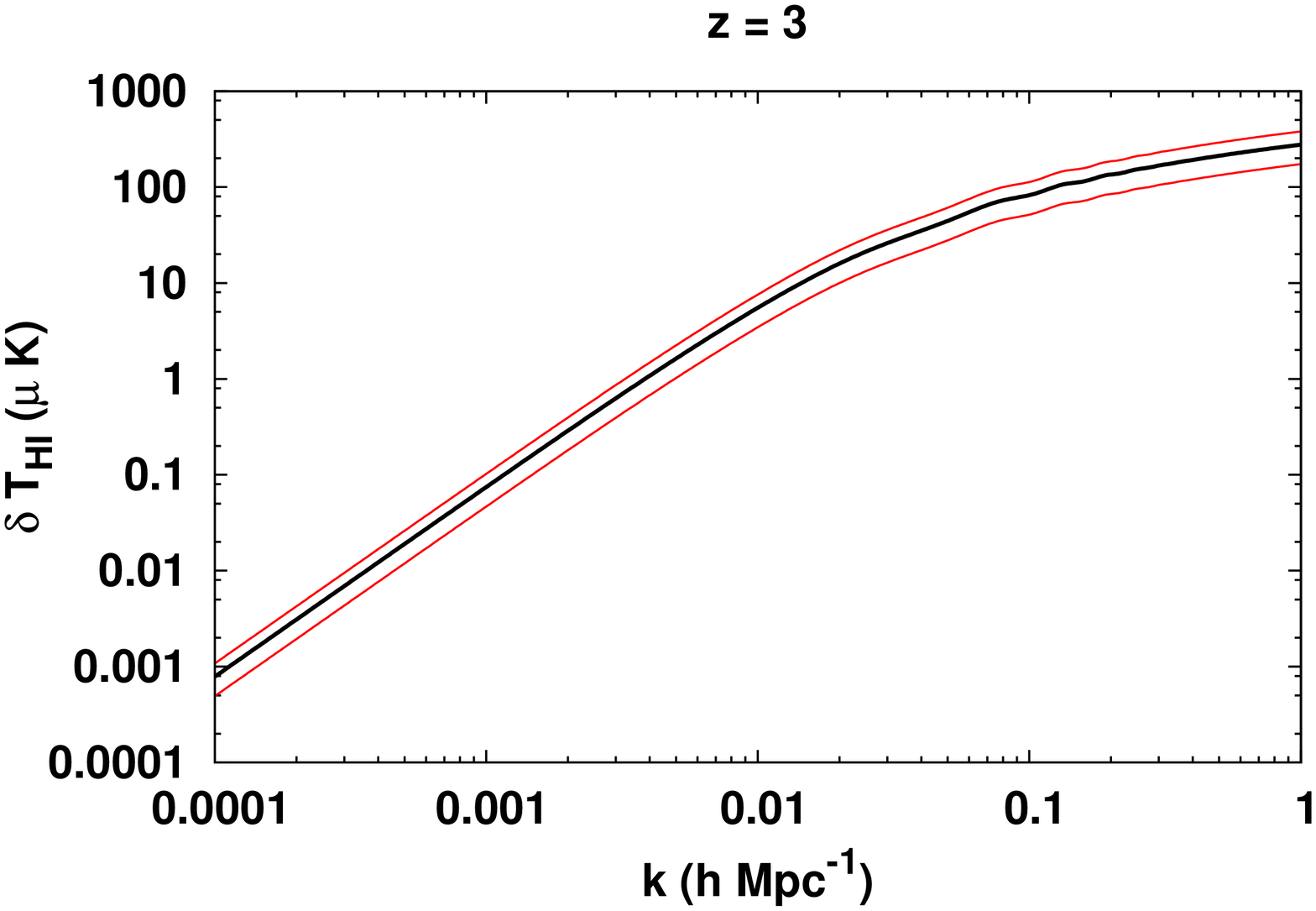}
  \end{center}
\caption{Impact on the HI power spectrum, $\delta T_{\rm HI}(k)$ due to the uncertainty in $\Omega_{\rm HI}$ and $b_{\rm HI}$ coming from the available measurements. Plots at redshifts 0, 1, 2 and 3 are shown.}
\label{fig:powspec}
 \end{figure}

 The above uncertainty on the power spectrum impacts the measurements by current and future intensity mapping experiments. To provide an indication of the significance of this effect, we
consider the expression for the signal-to-noise ratio (SNR) in the 21-cm signal \citep{feldman1994, seo2010, battye2012} for a single-dish radio experiment:
 \begin{equation}
\frac{S}{N} = \sqrt{\frac{2 \pi k^2 \Delta k V_{\rm sur}}{(2 \pi)^3}} \frac{P_{\rm HI}}{P_{\rm HI} + (\sigma_{\rm pix}^2 V_{\rm pix}/(\bar{T}(z)^2 \hat{W}(k)^2) + P_{\rm shot}}                                                                                                            \end{equation} 
In the above expression, $\Delta k$ is the wavenumber range and $V_{\rm sur}$ is the survey volume. $P_{\rm HI} \equiv (\delta T_{\rm HI})^2$ is the 3D power spectrum defined in \eq{3dpowspec}, $\bar{T}$ is the mean brightness temperature defined in \eq{tbar} and $P_{\rm shot}$ is the shot noise. The  $\sigma_{\rm pix}$ is the pixel noise defined by:
\begin{equation}
 \sigma_{\rm pix} = \frac{T_{\rm sys}}{\sqrt{t_{\rm pix} \delta \nu}}
\end{equation} 
where $T_{\rm sys}$ is the system temperature including both the instrument and the sky temperature, $t_{\rm pix}$ is the observation time per pixel and $\delta \nu$ is the frequency interval of integration. The window function $\hat{W}(k)$ models the angular and frequency response function of the instrument. Foreground removal may be contained in a residual noise term that remains after the foreground is assumed to be subtracted.

The above SNR, thus,  contains contributions from a noise term and a cosmic-variance term. If the intensity mapping experiment is noise-dominated, the noise term $(\sigma_{\rm pix}^2 V_{\rm pix}/(\bar{T}(z)^2 \hat{W}(k)^2) + P_{\rm shot}$ dominates $P_{\rm HI}$. In this case, the signal-to-noise ratio becomes proportional to the signal $P_{\rm HI}$. This indicates that the uncertainty in the signal translates into the uncertainty in the SNR. Hence, the observational uncertainties in the parameters $\Omega_{\rm HI}$ and $b_{\rm HI}$ have direct implications for the range of the signal-to-noise ratios of these experiments at different redshifts. In particular, the uncertainty of 50\% to 100\% (from Table \ref{table:uncert}) in the magnitude of the power spectrum $(\delta T_{\rm HI})^2$, implies the corresponding uncertainty in the signal-to-noise ratio. 

At large scales,  high-$\sigma$ detections with upcoming telescopes like the LOFAR and the SKA may be cosmic-variance dominated \citep[e.g.,][]{mesinger2014}. In these cases, the SNR is independent of the signal.

 \section{Conclusions}
 In this paper, we have considered recent available constraints on $\Omega_{\rm HI}$ and $b_{\rm HI}$ together with their allowed uncertainties, coming from a range of theoretical and observational sources. These are used to predict the consequent uncertainty in the HI power spectrum measured and to be measured by current and future experiments. Using a minimum variance interpolation scheme, we find that a combination of the available constraints allow a near 50\% - 100\% error in the measurement of the HI signal in the redshift range $z \sim 0 - 3.5$. This is essential for the planning and construction of the intensity mapping experiments. Table \ref{table:uncert} is of practical utility for quantifying the uncertainties in the various parameters.  We have tested three different confidence scenarios: optimistic, conservative and an intermediate scenario, and find the predicted uncertainties in all three cases to be fairly consistent over their range of applicability. It is also clear from the analysis that a constant value of either $\Omega_{\rm HI}$ or of $b_{\rm HI}$ does not fully take into account the magnitude of the uncertainties concerned. Hence, it is important to take into account the available measurements for a more precise prediction of the impact on the HI power spectrum.
 
 Even though we have assumed a standard $\Lambda CDM$ model for the purposes of this paper, the analysis may be reversed to obtain predictions for the cosmology, the evolution of the dark energy equation of state, curvature and other parameters \citep{chang10, bull2014}. Again, for such purposes, a realistic estimate of the input parameters ($\Omega_{\rm HI}, b_{\rm HI}$) would be useful to accurately predict the consequent uncertainties in the parameters predicted. 
 A model which accurately explains the value of bias at all redshifts, and the neutral hydrogen fraction is currently lacking and hence we use the present observations and theoretical prescriptions to provide the latest constraints on the 3D HI power spectrum. In the future, as better and more accurate measurements of the bias and neutral hydrogen density
 become available, it would significantly tighten our constraints on the power spectrum. Similarly, the clustering properties of DLAs which leads to the bias of DLAs at higher redshifts offers an estimate of the bias parameter of neutral hydrogen, though it is significantly higher. 
  
 We have indicated the implications of the predicted uncertainty in the power spectrum for the current and future intensity mapping experiments. In the case of a single-dish radio telescope, for example, the uncertainty in the power spectrum translates into an uncertainty in the signal-to-noise ratio (SNR) of the instrument in noise-dominated experiments. Thus, this has important consequences for the planning of HI intensity mapping measurements by current and future radio experiments.

 \section{Acknowledgements}
The research of HP is supported by the Shyama Prasad Mukherjee (SPM) research grant of the Council for Scientific and Industrial Research (CSIR), India. We thank Sebastian Seehars, Adam Amara, Christian Monstein, Aseem Paranjape and R. Srianand for useful discussions, and Danail Obreschkow, Jonathan Pober, Marta Silva and Stuart Wyithe for comments on the manuscript. HP thanks the Institute for Astronomy, ETH, Z\"{u}rich for hospitality during a visit when part of this work was completed. We thank the anonymous referee for helpful comments that improved the content and presentation.

\bibliographystyle{mn2e} 
\def\aj{AJ}                   
\def\araa{ARA\&A}             
\def\apj{ApJ}                 
\def\apjl{ApJ}                
\def\apjs{ApJS}               
\def\ao{Appl.Optics}          
\def\apss{Ap\&SS}             
\def\aap{A\&A}                
\def\aapr{A\&A~Rev.}          
\def\aaps{A\&AS}              
\def\azh{AZh}                 
\def\baas{BAAS}
\def\jcap{JCAP}
\def\jrasc{JRASC}             
\def\memras{MmRAS}
\def\na{New Astronomy}
\def\nat{Nature}
\def\mnras{MNRAS}             
\def\pra{Phys.Rev.A}          
\def\prb{Phys.Rev.B}          
\def\prc{Phys.Rev.C}          
\def\prd{Phys.Rev.D}          
\def\prl{Phys.Rev.Lett}       
\def\pasp{PASP}               
\def\pasj{PASJ}
\def\physrep{Phys. Repts.}
\def\qjras{QJRAS}             
\def\skytel{S\&T}             
\def\solphys{Solar~Phys.}     
\def\sovast{Soviet~Ast.}      
\def\ssr{Space~Sci.Rev.}      
\def\zap{ZAp}                 
\let\astap=\aap
\let\apjlett=\apjl
\let\apjsupp=\apjs

\bibliography{mybib}

\appendix

\section{Conservative and optimistic estimates on the uncertainties in the HI power spectrum}

In this appendix, we consider the two additional possible scenarios of modelling the uncertainties on the bias parameter $b_{\rm HI}$, which were denoted by cases (a) and (b) in Section \ref{sec:combine} of the main text. 

(a) Conservative:  This approach has the justification that it utilizes all the available observations and their associated error bars, and avoids any ambiguity related with assigning errors to simulation data. However, since the observations are limited to $z \lesssim 1$, the minimum variance estimator is also limited to this redshift range, with associated uncertainties that use only the two available $b_{\rm HI}$ measurements at $z \lesssim 1$. This is plotted in Fig. \ref{fig:powspec_conserv} along with the estimate for the product $\Omega_{\rm HI} b_{\rm HI}$, and the table of predicted uncertainties is provided in Table \ref{table:conservative}. Over the relevant redshift range $z \lesssim 1$, the constraints are fairly similar to those in the intermediate scenario (considered in the main text). Since we only have two observational data points over this redshift range, the mean values and uncertainties depend only upon these two observational measurements. Hence, the constraints on the bias $b_{\rm HI}$ are also expected to be of the same order as those in the intermediate scenario, over this redshift range. However, we cannot predict uncertainties in the bias and the power spectrum for redshifts $z > 1$ due to the unavailability of observational data, and hence this scenario may be termed conservative.

\begin{figure}
  \begin{center}
   \includegraphics[scale = 0.32, angle = -90]{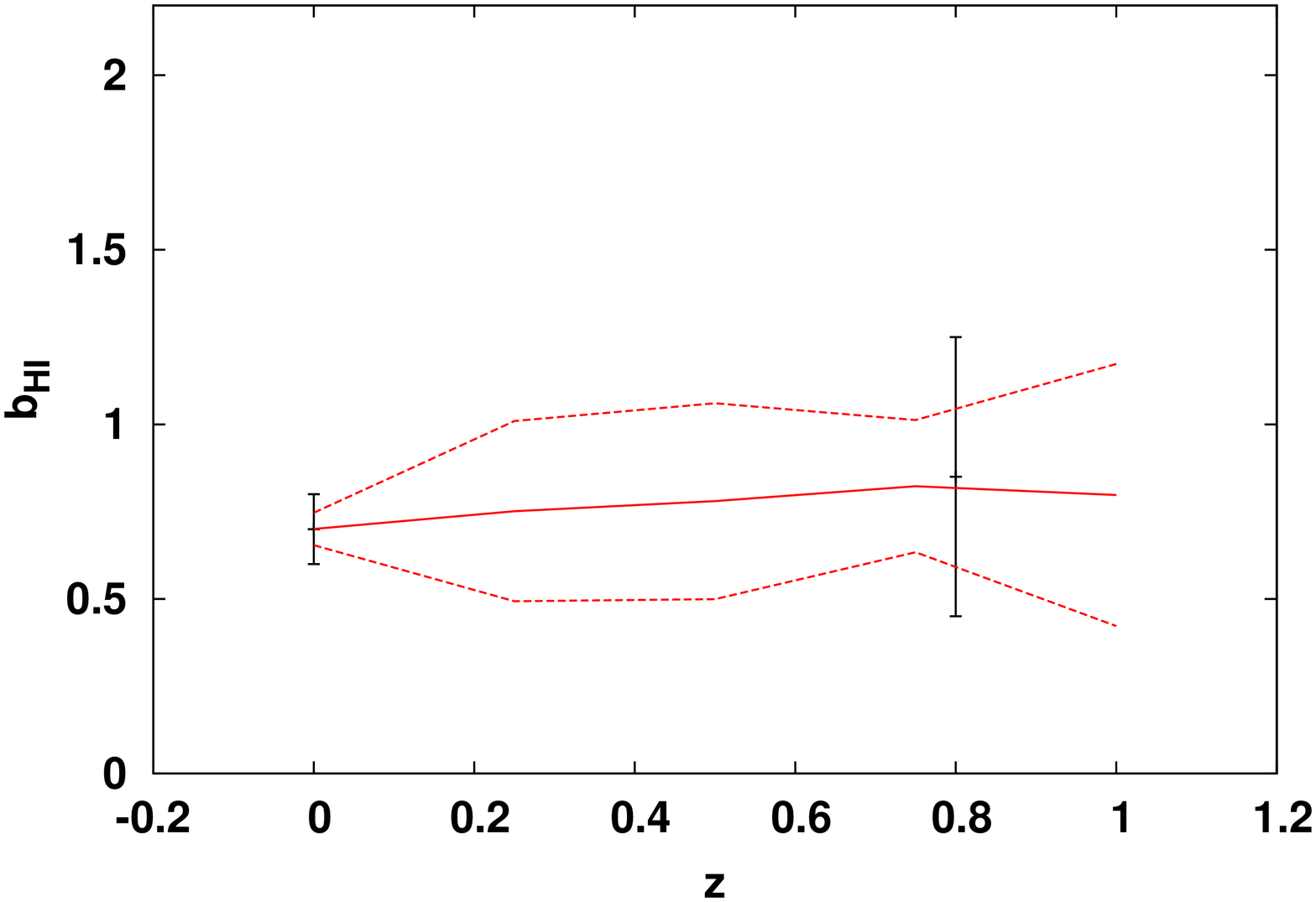} \includegraphics[scale = 0.32, angle = -90]{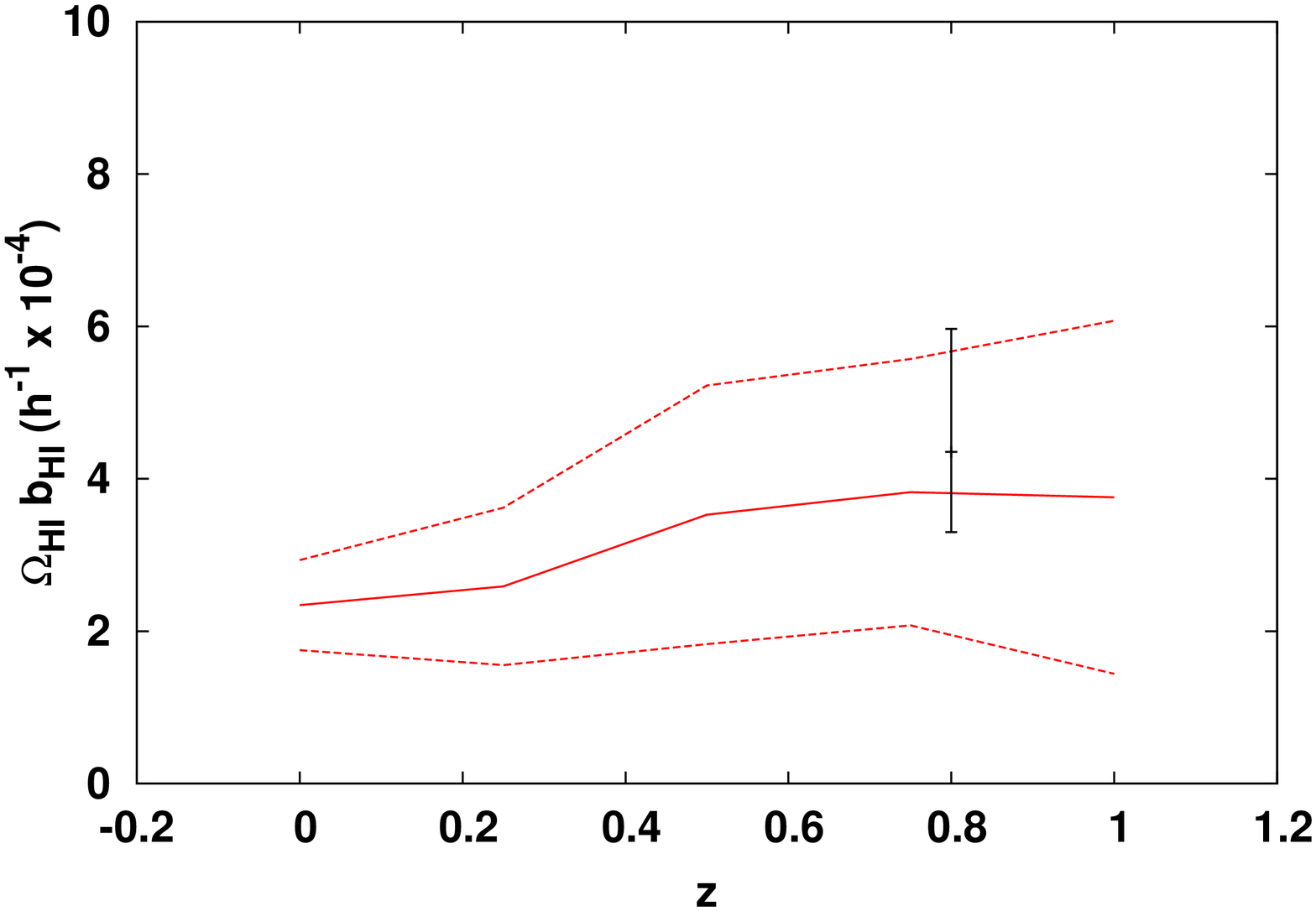} 
  \end{center}
\caption{Conservative estimates for $b_{\rm HI}$ and the product $\Omega_{\rm HI} b_{\rm HI}$, taking into account the available observations only, without any theoretical predictions. The measurement \citep{switzer13} at $z = 0.8$ is overplotted on the product curve for reference.}
\label{fig:powspec_conserv}
 \end{figure}
 
 \begin{table*}
 \begin{center}
   \hspace{0in}  \begin{tabular}{lcllllll}
    \hline
    $z$   & $\Omega_{\rm HI} ^{\dagger}$ & $\Delta \Omega_{\rm HI}^{\dagger}$ & $b_{\rm HI}$ & $\Delta b_{\rm HI}$ & $\Omega_{\rm HI} b_{\rm HI}^{\dagger}$ & $\Delta (\Omega_{\rm HI}  b_{\rm HI})^{\dagger}$ & $\Delta (\Omega_{\rm HI}  b_{\rm HI})/(\Omega_{\rm HI} b_{\rm HI})$\\ 
   \hline 
   0.000         &   3.344         &   0.814         &   0.700         &   0.046         &   2.342         &   0.591         &   0.252        \\
   0.250         &   3.443         &   0.703         &   0.751         &   0.258         &   2.587         &   1.033         &   0.399        \\
   0.500         &   4.523         &   1.445         &   0.780         &   0.280         &   3.527         &   1.696         &   0.481        \\
   0.750         &   4.648         &   1.835         &   0.823         &   0.189         &   3.825         &   1.748         &   0.457        \\
   1.000         &   4.710         &   1.877         &   0.798         &   0.376         &   3.757         &   2.317         &   0.617        \\
   \hline 
   $\dagger$ In units of $10^{-4} h^{-1}$. 
   \end{tabular}   
   \end{center}
 \caption{Same as Table \ref{table:uncert} for the ``conservative'' case where only observational uncertainties contribute to $b_{\rm HI}$. Note that $\Omega_{\rm HI}$ is in units of $10^{-4} h^{-1}$.}
\label{table:conservative}
 \end{table*}
 
(b) Optimistic: Motivation for this approach comes from providing a strict lower limit to the uncertainties in the HI signal, using the uncertainties in $\Omega_{\rm HI}$ alone. Here, we consider a theoretical model\footnote{We emphasize that the model under consideration is only for illustrative purposes, since our aim is to quantify the uncertainty in the HI signal rather than to forecast the magnitude of the signal.} which predicts the value of $b_{\rm HI}$ at all redshifts \citep{bagla2010}. We combine the predictions  of the bias from the model, assuming negligible errors, with the observational constraints on $\Omega_{\rm HI}$. Fig. \ref{fig:snakeforprod_optim} shows the resulting uncertainty on the product $\Omega_{\rm HI} b_{\rm HI}$, and Table \ref{table:optimistic} tabulates the uncertainties.  We note that this scenario, while being optimistic, (a) uses both the theoretical (for the mean value) and observational (for the error bars) constraints on the parameters $b_{\rm HI}$ and $\Omega_{\rm HI}$ respectively, and, (b) importantly,  recovers a lower limit on the predicted HI uncertainty.

\begin{figure}
  \begin{center}
   \includegraphics[scale = 0.32, angle = -90]{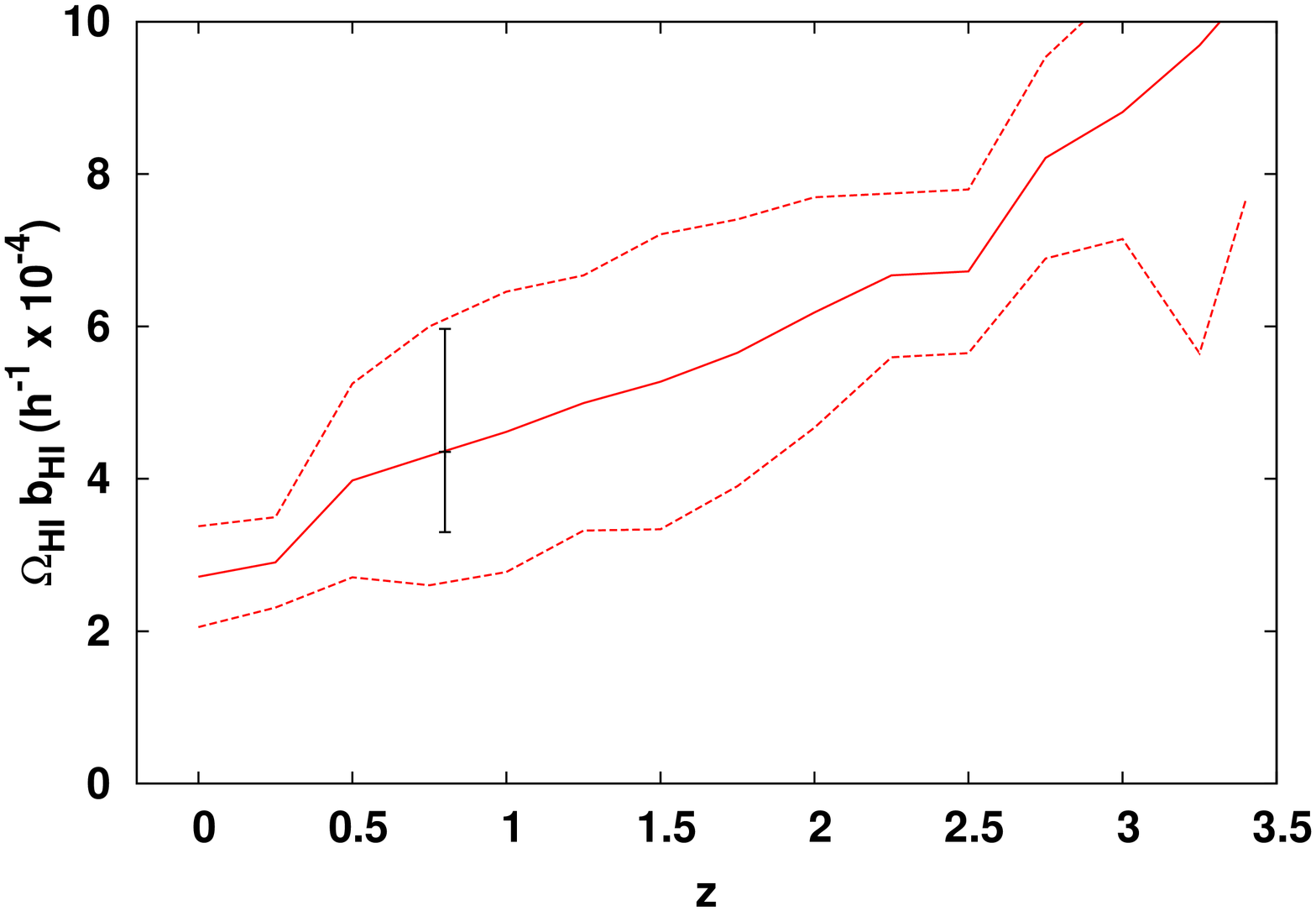} 
  \end{center}
\caption{Optimistic estimates for the product $\Omega_{\rm HI} b_{\rm HI}$, taking into account the available observations for the uncertainties in $\Omega_{\rm HI}$ and neglecting uncertainties in $b_{\rm HI}$ associated with theory/simulations. The measurement \citep{switzer13} at $z = 0.8$ is also overplotted for reference.}
\label{fig:snakeforprod_optim}
 \end{figure}

\begin{table*}
 \begin{center}
   \hspace{0in}  \begin{tabular}{lcllllll}
    \hline
    $z$   & $\Omega_{\rm HI} ^{\dagger}$ & $\Delta \Omega_{\rm HI}^{\dagger}$ & $b_{\rm HI}$ & $\Omega_{\rm HI} b_{\rm HI}^{\dagger}$ & $\Delta (\Omega_{\rm HI}  b_{\rm HI})^{\dagger}$ & $\Delta (\Omega_{\rm HI}  b_{\rm HI})/(\Omega_{\rm HI} b_{\rm HI})$\\ 
   \hline 
        0.000         &   3.344         &   0.814         &   0.812         &   2.715         &   0.661         &   0.243        \\
   0.250         &   3.443         &   0.703         &   0.843         &   2.903         &   0.592         &   0.204        \\
   0.500         &   4.523         &   1.445         &   0.880         &   3.978         &   1.271         &   0.319        \\
   0.750         &   4.648         &   1.835         &   0.925         &   4.301         &   1.698         &   0.395        \\
   1.000         &   4.710         &   1.877         &   0.980         &   4.616         &   1.839         &   0.398        \\
   1.250         &   4.804         &   1.612         &   1.040         &   4.995         &   1.676         &   0.336        \\
   1.500         &   4.766         &   1.750         &   1.106         &   5.273         &   1.936         &   0.367        \\
   1.750         &   4.804         &   1.487         &   1.177         &   5.654         &   1.750         &   0.310        \\
   2.000         &   4.936         &   1.207         &   1.253         &   6.184         &   1.512         &   0.245        \\
   2.250         &   5.008         &   0.807         &   1.332         &   6.669         &   1.075         &   0.161        \\
   2.500         &   4.750         &   0.759         &   1.415         &   6.722         &   1.074         &   0.160        \\
   2.750         &   5.471         &   0.880         &   1.501         &   8.213         &   1.321         &   0.161        \\
   3.000         &   5.541         &   1.048         &   1.591         &   8.814         &   1.668         &   0.189        \\
   3.250         &   5.756         &   2.401         &   1.683         &   9.689         &   4.042         &   0.417        \\
   3.400         &   5.971         &   1.570         &   1.739         &  10.386         &   2.730         &   0.263        \\
   \hline 
   $\dagger$ In units of $10^{-4} h^{-1}$. 
   \end{tabular}   
   \end{center}
 \caption{Same as Table \ref{table:uncert} for the ``optimistic'' case, where only uncertainties in $\Omega_{\rm HI}$ are considered, assuming that $\Delta b_{\rm HI} = 0$ for all redshifts. The final column provides strict lower limits on the relative uncertainty in the amplitude of the HI signal. Note that $\Omega_{\rm HI}$ is in units of $10^{-4} h^{-1}$.}
\label{table:optimistic}
 \end{table*}

\end{document}